\def\doublecolumn{1} 
\if 1\doublecolumn
  \documentclass[journal]{IEEEtran}
\else
  \documentclass[12pt, draftclsnofoot, onecolumn, letterpaper]{IEEEtran}
\fi

\def\blind{1} 

\usepackage{custom_package}
\usepackage{enumitem}
\setlist{leftmargin=3.5mm}

\newcommand{\MaskCode}{{\textsc{MaskCode}}}

\begin{document}

\title{
    \MaskCode: Mask Transformer for Feedback-Assisted Coding With Linear Block Codes
}

\if 1\blind
\author{
    Jonggyu Jang, 
    Hongjae Nam,
    Vishrant Tripathi, 
    David J. Love,
    and Hyun Jong Yang
    \thanks{
    J. Jang is with the Department of Electronics Engineering, Chungnam National University, Daejeon 34134, Republic of Korea (e-mail: jgjang@cnu.ac.kr).
    H. Nam, V. Tripathi, and D. J. Love are with the Elmore Family School of Electrical and Computer Engineering, Purdue University, West Lafayette, IN 47907 USA (e-mail: \{nam86,  tripathv, djlove\}@purdue.edu).
    H. J. Yang is with the Department of Electrical and Computer Engineering, Seoul National University, Seoul 08826, Republic of Korea (e-mail: hjyang@snu.ac.kr).
    The corresponding author is David J. Love. 
    This work was supported in part by the Office of Naval Research (ONR) under Grant N000142112472, in part by the National Science Foundation (NSF) under Grants CNS2212565, CNS2225578, and EEC1941529, and in part by the National Research Foundation of Korea(NRF) grant funded by the Korea government(MSIT) (RS-2026-25490601).
    }
}
\else
\author{Anonymous Submission
    }
\fi
\maketitle

\begin{abstract}
    Feedback-based coding schemes have demonstrated substantial performance gains over today's open-loop coding schemes.  
    Unfortunately, these gains are usually achieved in idealized settings with perfect feedback.  
    Over the last few years, machine learning-based schemes have been shown to be promising solutions for implementing feedback-based codes, particularly when combined with short-block-length open-loop error correcting codes (ECCs) in a concatenated coding structure.
    However, existing ML-based feedback schemes remain agnostic to the outer code's structure, potentially misallocating feedback resources on error patterns already correctable by the outer ECC. To address this, we propose \MaskCode, a Transformer-based inner feedback code for concatenated coding systems, which explicitly incorporates structural knowledge of the outer linear block code into the inner feedback encoder design via two synergistic mechanisms: 1) \textit{a soft syndrome-based input} that informs the encoder about potential parity constraint violations, and 2) \textit{a code-aware attention mask} derived from the Tanner graph.
    We further show that end-to-end training with a differentiable belief propagation (BP) decoder offers no additional gain, as \MaskCode's structure-aware design already internalizes the structural knowledge of the outer code; in fact, backpropagation through the iterative BP decoder introduces gradient explosion, which degrades rather than improves performance.
    Extensive evaluations on BCH and LDPC outer codes demonstrate that \MaskCode~consistently outperforms all baselines, achieving up to 1.5 dB SNR gain.
\end{abstract}

\section{Introduction}

Over the past decade, artificial intelligence and machine learning (AI/ML) have emerged as a potential key enabler of 6G communications~\cite{brinton2025key}.
More specifically, deep learning (DL) leverages deep neural networks (DNNs) trained via data-driven optimization to represent intricate nonlinear functions.
By virtue of the data-driven optimization, AI/ML has effectively addressed mathematically intractable problems across a wide range of telecommunication technologies, such as channel prediction, channel decoding, radio resource allocation, network optimization, and waveform learning~\cite{kim2025machine,9207875,choukroun2022error,10604898,11017513,park2025lowering,ahmed2019deep}.

One notable application of ML-based methods in wireless communications is \textit{feedback coding}. 
In early work, Shannon showed that feedback does not increase channel capacity for memoryless channels~\cite{1056798, kadota1971capacity}; however, subsequent studies~\cite{sk1966coding-1,sk1966coding-2, 10078921, burnashev1976data} have shown that closed-loop feedback codes enable the probability of error to decay \textit{double-exponentially} with blocklength, by adaptively refining the transmission using receiver-side information.
An early linear realization, the Schalkwijk-Kailath (SK) scheme~\cite{sk1966coding-1,sk1966coding-2}, achieves capacity and exhibits an error probability that decays double-exponentially.
When the feedback link is itself noisy, however, the SK scheme loses not only the double-exponential decay but also the capacity-achieving rate~\cite{kim2006feedback, kim2007gaussian, kim2009feedback}.
In this noisy-feedback regime, the Chance-Love (CL) scheme~\cite{chance2011concatenated} achieves the error exponent bound established in~\cite{kim2007gaussian}, but its encoding and decoding structures remain constrained to linear operations.
More recently, ML-based feedback coding~\cite{kim2018deepcode,pmlr-v202-kim23j,ozfatura2022all,shao2023attentioncode,ankireddy2025lightcode} replaces the linear encoder and decoder with DNNs, demonstrating substantial performance gain over linear schemes.

\begin{figure}[t]
    \centering
    \includegraphics[width=0.95\linewidth]{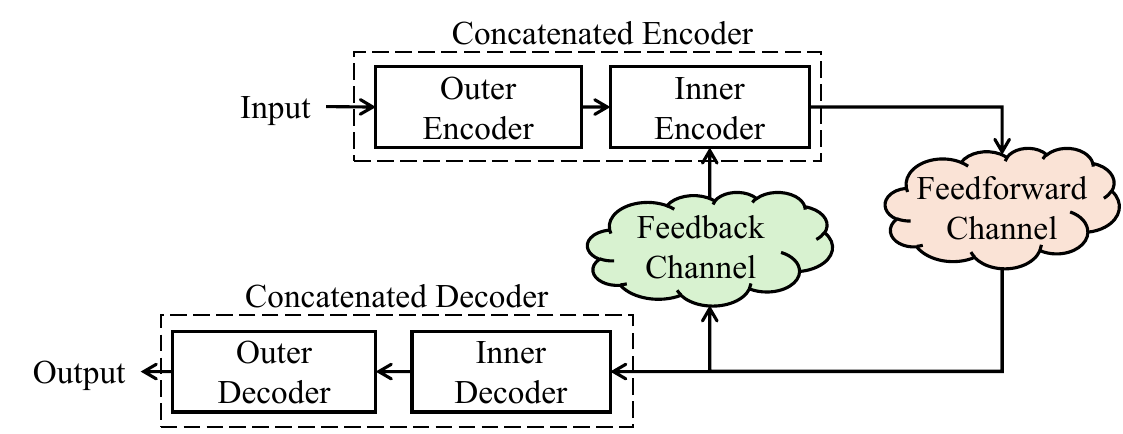}
    \caption{Illustration of the feedback-assisted coding system with an outer linear block code and an inner feedback code.}
    \label{fig:concatenated_concept_sketch}
\end{figure}

Beyond standalone feedback coding, \textit{concatenated coding} offers a promising research direction that integrates feedback coding with conventional error-correcting codes (ECCs)~\cite{forney1965concatenated, wicker1999reed, cain1970concatenation, chance2011concatenated}. 
These coding architectures employ a two-stage structure consisting of an inner code and an outer code, as depicted in \cref{fig:concatenated_concept_sketch}, where the outer code is a general open-loop ECC, and the inner code is a feedback coding scheme that exploits closed-loop communication channels.
This concatenated design raises the question of how the inner feedback code should be designed to best support the outer decoder.

\subsection{Related Works}

Here, we provide a literature survey for 1) linear feedback coding, 2) analytical nonlinear feedback coding, 3) ML-based feedback coding, 4) feedback coding in concatenated coding architecture, and 5) ML-based ECC decoders.

\paragraph*{Linear feedback coding}
As a seminal work, Schalkwijk \textit{et al.} proposed a capacity-achieving linear feedback coding strategy for Gaussian channels with the noiseless feedback setting, which we refer to as the SK scheme~\cite{sk1966coding-1,sk1966coding-2}.
However, once the feedback becomes noisy, the SK scheme loses both its capacity-achieving property and its double-exponential error decay. To address this limitation, several linear feedback coding schemes have been proposed based on analytical optimization, notably the CL scheme~\cite{chance2011concatenated} and dynamic programming~\cite{10078921}.
In~\cite{agrawal2011iteratively,7274698}, extensions of the CL scheme were developed for correlated noise and broadcasting channel scenarios.
However, these schemes are restricted to linear encoding and decoding structures. As a result, they do not capture the potential gains of more general nonlinear feedback schemes.

\paragraph*{Analytical nonlinear feedback coding}
To circumvent the performance bottlenecks imposed by linear schemes, several studies have explored the integration of nonlinear operations into feedback coding~\cite{gallager2009variations,ben2014gaussian,7809083,ankireddy2025lightcode}.
In~\cite{gallager2009variations}, Gallager-Nakibo{\u{g}}lu (GN) scheme exploits nonlinearities by correcting single-distance errors via pulse amplitude modulation (PAM) signaling. 
On the other hand, in~\cite{ben2014gaussian,7809083}, the Modulo-SK scheme was proposed to handle noisy feedback by leveraging a nonlinear modulo operation.
Most recently, PowerBlast~\cite{ankireddy2025lightcode} was proposed as a hybrid architecture, combining the SK scheme and GN scheme.
All the aforementioned studies have proposed nonlinear operations; however, these nonlinear operations (\eg, modulo and PAM signaling) are manually engineered, thereby lacking flexible representation.

\paragraph*{ML-based feedback coding}
Recently, DNN-based feedback coding has emerged as a compelling alternative to linear schemes, unlocking the potential performance gains of nonlinear feedback coding. 
In~\cite{kim2018deepcode}, a pioneering scheme, DeepCode, was developed, which uses recurrent neural network (RNN) architectures for both encoder and decoder. 
Even with only two RNN layers, DeepCode can outperform linear methods, such as SK scheme~\cite{sk1966coding-1,sk1966coding-2} and CL scheme~\cite{chance2011concatenated}, in certain regimes in terms of BLER.
Following the success of AI/ML in feedback coding, subsequent works have explored various architectures, including RNNs~\cite{pmlr-v202-kim23j, safavi2021deep, mashhadi2021drf}, Transformers~\cite{shao2023attentioncode,ozfatura2022all}, and light-weight neural networks~\cite{ankireddy2025lightcode}, and two-way learning-based extensions comparing multiple neural architectures~\cite{kim2025coding, nickel2025learning}.

\paragraph*{Feedback coding in concatenated codes}
The concept of concatenated coding was first introduced by Forney~\cite{forney1965concatenated}, which has been historically validated in a wide variety of deployments, including NASA's deep space mission~\cite{wicker1999reed}. 
In concatenated coding frameworks, feedback coding can serve as the inner code.
The authors of~\cite{cain1970concatenation} theoretically showed that the use of feedback yields superior error exponents compared to those achievable without feedback.
In~\cite{chance2011concatenated}, the authors introduced the CL scheme as a feedback-driven inner code for outer ECCs, thereby exploiting the feedback gain in concatenated coding.
In~\cite{kim2018deepcode}, DeepCode was shown to outperform the CL scheme and SK scheme when used as an inner code with an outer Turbo code.

\paragraph*{ML-based ECCs}

As another application of AI/ML, ECC decoding has been extensively investigated using DNNs~\cite{choukroun2022error, 10960678, park2024crossmpt, park2025lowering}. 
A prominent trend in this line of research is the adoption of Transformer architectures. 
A representative work, ECCT~\cite{choukroun2022error}, and its extensions~\cite{park2025lowering,10960678} employ attention masks derived from hard-decision syndrome information to steer the model toward code-structure-aware representations, thereby enhancing decoding performance.

\subsection{Challenges and Contributions}
While ML-based feedback coding has shown substantial gains over linear schemes, existing studies have largely focused on the standalone setting~\cite{kim2018deepcode,pmlr-v202-kim23j,ozfatura2022all,shao2023attentioncode,ankireddy2025lightcode}, leaving its design within concatenated architectures underexplored.
Because the inner feedback code determines the quality of the effective channel (\textit{super-channel})~\cite{forney1965concatenated} seen by the outer decoder, the structural limitations imposed by linear inner feedback codes carry over directly to concatenated systems, motivating the use of ML-based nonlinear inner feedback codes.
Although~\cite{kim2018deepcode} has shown the promise of this direction by combining an ML-based inner code with an outer Turbo code~\cite{397441}, no comprehensive framework exists for the synergistic co-design of ML-based inner feedback codes and outer decoders.
To bridge this research gap, this paper identifies and addresses the following challenges:
\begin{itemize} 
    \item \textbf{Challenge 1: Lack of synergistic design. } 
    When employed as inner codes in concatenated codes, existing ML-based feedback coding schemes---trained solely to minimize their own bit-error rate (BER) by treating the input bits as independent---ignore the algebraic constraints of the outer code~\cite{kim2018deepcode,pmlr-v202-kim23j,ozfatura2022all,shao2023attentioncode,ankireddy2025lightcode}, \ie, parity checking, $\mathbf{cH}^\mathrm{T}=\mathbf{0}$ for codeword $\mathbf{c}$ and parity check matrix $\mathbf{H}$.
    Hence, while compatible, they are ill-suited to the structure of outer ECCs, leading to suboptimal performance in concatenated systems.
    \item \textbf{Challenge 2: End-to-end training. } End-to-end training strategies, which jointly optimize the inner code with the outer decoder, serve as a direct and intuitive way for synergistic design.
    While such paradigms have been proposed for Turbo and convolutional codes~\cite{jiang2019turbo,streit2025transformer}, the efficacy of end-to-end training with inner feedback coding remains unexplored---a non-trivial problem, as incorporating a differentiable iterative outer decoder introduces gradient explosion during backpropagation.
\end{itemize}

\paragraph*{Contribution}
To address these challenges, we propose \MaskCode, a Transformer-based architecture where the inner encoder and decoder are designed to internalize the \textit{structural design of the outer ECCs}.
Specifically, we consider two representative linear block codes as outer ECCs: Bose-Chaudhuri-Hocquenghem (BCH)~\cite{hocquenghem1959codes,bose1960class} and low-density parity-check (LDPC)~\cite{gallager1962low,mackay1996near}, both of which can be represented as Tanner graphs and decoded via belief propagation (BP), enabling differentiable end-to-end training.
In a nutshell, our contributions are threefold:
\begin{itemize}
    \item \textbf{Synergistic inner feedback coding: } In \MaskCode, we propose two key mechanisms to internalize the structural constraints of the outer code: 1) a \textit{soft-syndrome-based input} that enables the inner code to recognize parity-check constraints, and 2) a \textit{masking strategy} that repurposes the self-attention mask to enforce code-structure-aware attention patterns derived from the Tanner graph, ensuring only valid message-passing interactions are learned.
    By doing so, \MaskCode~strengthens bits that are difficult for the outer decoder to correct.
    \item \textbf{Analysis of end-to-end training: } 
    We numerically evaluate the effect of end-to-end training on feedback-assisted concatenated codes by comparing ML-based models trained with various numbers of BP iterations. 
    Counterintuitively, three or more BP iterations degrade BLER performance, which we attribute to the gradient instability introduced by unrolling BP iterations.
    In particular, we show that the integration of the outer code structure into the \MaskCode~architecture makes \textit{end-to-end training unnecessary}.
    \item \textbf{Comprehensive benchmark results for concatenated code systems:} 
    Unlike previous studies~\cite{ankireddy2025lightcode,shao2023attentioncode,ozfatura2022all,pmlr-v202-kim23j}, we provide a systematic benchmark evaluating ML-based feedback coding schemes as inner codes in concatenated code systems.
\end{itemize}

\paragraph*{Notations}
Throughout this paper, we will use the following notations.
Lowercase boldface letters represent \textbf{\textit{row}} vectors, \eg, $\mathbf{v} \in \mathbb{R}^{1\times n}$.
The $i$-th element of the vector $\mathbf{v}$ is written as $v_i$.
Also, for vectors with a subscript, the $i$-th element of the vector $\mathbf{v}_k$ is written as $v_{k,i}$.
The ($i,j$)-th element of matrix $\mathbf{M}$ is denoted as $[\mathbf{M}]_{i,j}$.
For an integer $K$, $[K]$ represents the set of integers from $1$ to $K$, \ie, $[K] = \{1, 2, \dots, K\}$.

\section{Background and Preliminaries}

To facilitate a better understanding of the proposed framework, this section establishes the necessary background regarding ECCs and Transformer models~\cite{vaswani2017attention}.
Experts and practitioners with a pre-established understanding of ECCs and the Transformer model may choose to skip this section.

\subsection{Error Correcting Codes}
\label{subsec:ECCs}

In our concatenated coding system, we employ a linear block code as an outer code.
Let $\mathcal{C}$ be an $(n,k)$ binary linear block code over $\mathbb{F}_2:=\{0,1\}$, characterized by a code length $n$ and message length $k$ with the code rate of $\nicefrac{k}{n}$.
The linear code $\mathcal{C}\subseteq \{0,1\}^{1\times n}$ is described by a generator matrix $\mathbf{G} \in \{0,1\}^{k\times n}$ and a parity-check matrix $\mathbf{H} \in \{0,1\}^{(n-k) \times n}$, satisfying $\mathbf{G}\mathbf{H}^\mathrm{T}=\mathbf{0}_{k\times(n-k)}$ \cite{mceliece2002theory}. 
First, the encoder produces codeword $\mathbf{c}\in\{0,1\}^{1\times n}$ from the binary message $\mathbf{m}\in \{0,1\}^{1\times k}$ using a linear mapping defined by $\mathbf{G}$, \ie,
\begin{equation}\label{eq:cmG}
    \mathbf{c}=\mathbf{m}\mathbf{G} \in \{0,1\}^{1\times n}.
\end{equation}
In particular, the row space of $\mathbf{G}$ defines the linear code $\mathcal{C}$, and every codeword $\mathbf{c} \in \mathcal{C}$ is orthogonal to each row of $\mathbf{H}$, \ie,  $\mathbf{cH}^\mathrm{T}=\mathbf{0}$ for all $\mathbf{c} \in \mathcal{C}$.
Let $\mathbf{y}\in\mathbb{R}^{1\times n}$ denote the channel observation corresponding to the transmitted codeword $\mathbf{c}$. The hard-decision of $\mathbf{y}$ is denoted by $\mathbf{u}\in\{0,1\}^{1\times n}$.
The hard syndrome is defined as $\mathbf{uH}^\mathrm{T}\in\{0,1\}^{1\times (n-k)}$, which measures the parity-check violations and satisfies $\mathbf{uH}^\mathrm{T}=\mathbf{0}_{1\times (n-k)}$ if and only if $\mathbf{u}$ is a valid codeword.
The ECC decoder relies on the structural knowledge that $\mathbf{c}\mathbf{H}^\mathrm{T}=\mathbf{0}_{1\times(n-k)}$, regardless of the underlying channel medium, such as a raw physical channel or a virtual super-channel formed by an inner code.

\paragraph*{Belief propagation}
In this work, we adopt the sum-product algorithm (the most widely used BP algorithm for decoding linear block codes) as the outer decoder, referred to as the BP decoder.
The BP operates on a Tanner graph, a bipartite graph used to specify an ECC, consisting of $n$ variable nodes and $n-k$ check nodes, where an edge exists between variable node $v$ and check node $c$ if $[\mathbf{H}]_{c,v}=1$.
For the BP decoder, we define that the LLRs of the decoder's input $\widehat{\mathbf{c}}$, \ie
\begin{equation}
     L_i = \widehat{c}_i= \log \left(\frac{\Pr(c_i=1|\mathbf{y})}{\Pr(c_i=0|\mathbf{y})}\right).
\end{equation}
Then, the BP algorithm iteratively exchanges two types of messages over each edge of the Tanner graph, a variable-to-check message $\mu_{\text{vc}}^{i\rightarrow j}$ and a check-to-variable message $\mu_{\text{cv}}^{j\rightarrow i}$, initialized as $\mu_{\text{cv}}^{j\rightarrow i}=0$ for all edges~\cite{kschischang2001factor}:
\begin{enumerate}
    \item \textbf{Variable-to-check message ($\mu_{\text{vc}}^{i\rightarrow j}$):} The variable-to-check message aggregates beliefs from the neighboring check nodes by 
    \begin{equation}\label{eq:vtoc}
        \mu_{\text{vc}}^{i\rightarrow j} \leftarrow L_i + \sum_{k\in N_\text{c}(i)\setminus\{j\}} \mu_{\text{cv}}^{k\rightarrow i},
    \end{equation}
    where $N_\text{c}(i)=\{j\in[n-k]~|~[\mathbf{H}]_{j,i}=1\}$.
    \item \textbf{Check-to-variable message ($\mu_{\text{cv}}^{j\rightarrow i}$):} the check-to-variable message aggregates belief from the neighboring variable nodes by 
    \begin{equation}
        \mu_{\text{cv}}^{j\rightarrow i}\leftarrow 2\mathrm{arctanh}\left(\prod_{k\in N_\text{v}(j)\setminus i} \tanh\left(\frac{\mu_{\text{vc}}^{k\rightarrow j}}{2}\right)\right),
    \end{equation}
    where $N_\text{v}(j)=\{i\in[n]~|~[\mathbf{H}]_{j,i}=1\}$.
\end{enumerate}
After sufficient iterations, the final posterior LLR is computed as 
\begin{equation}
    L'_i =L_i + \sum_{k\in N_\text{c}(i)} \mu_{\text{cv}}^{k\rightarrow i},~\forall i\in[n],
\end{equation}
which is then used to reconstruct the message $\widehat{\mathbf{m}}$.
In the remainder of this work, the BP decoding algorithm is denoted as
\begin{equation}\label{eq:def_BP}
    \widehat{\mathbf{c}}_{\text{out}} = \mathtt{BP}(\widehat{\mathbf{c}}_{\text{in}}; N_\text{BP}),
\end{equation}
where $\widehat{\mathbf{c}}_{\text{in}} \in \mathbb{R}^{1 \times n}$ and $\widehat{\mathbf{c}}_{\text{out}} \in \mathbb{R}^{1 \times n}$ denote the input and output LLR vectors of the BP decoder, respectively, and $N_\text{BP}$ is the number of message-passing iterations.

\begin{figure}
    \centering
    \includegraphics[width=1.0\linewidth]{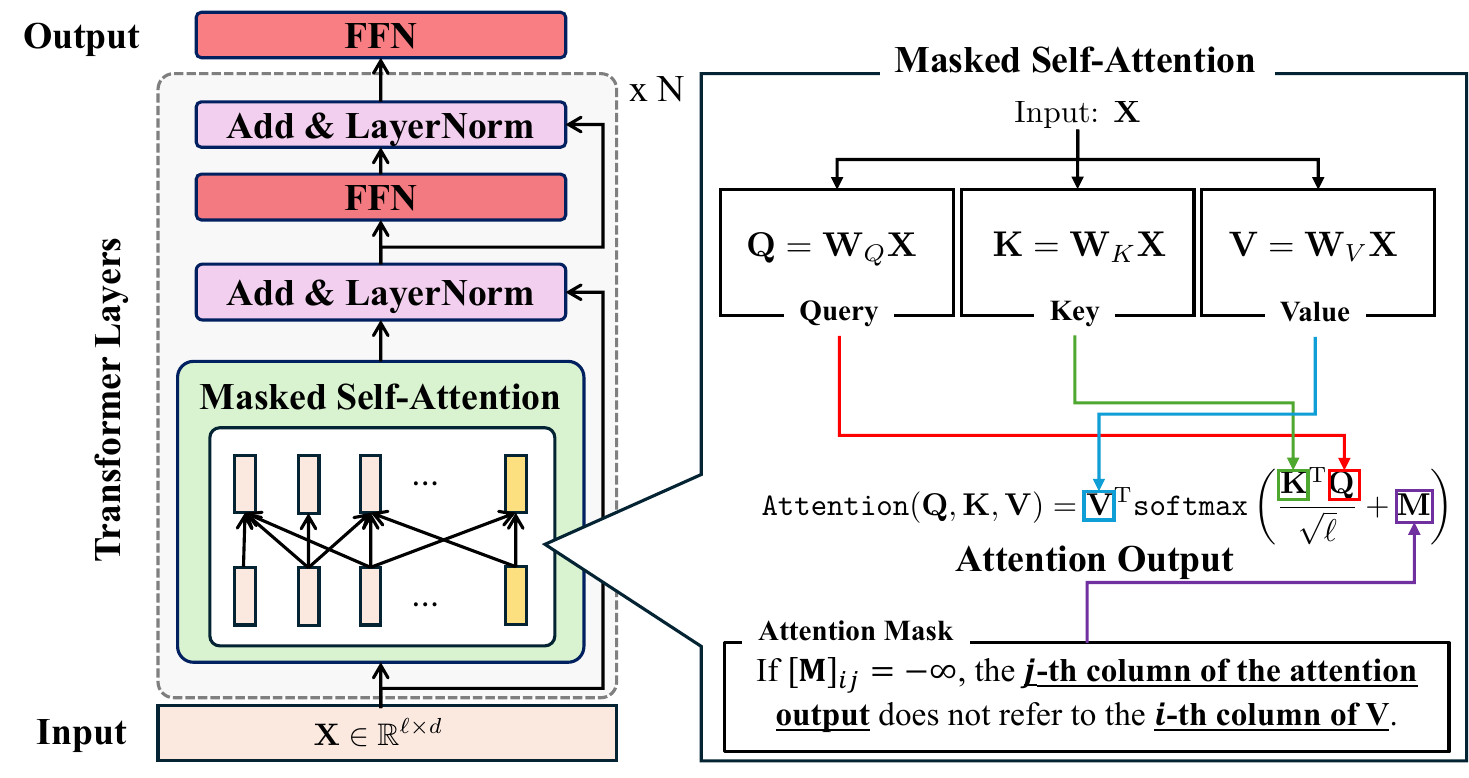}
    \caption{An illustration of a Transformer layer, consisting of 1)  a masked self-attention block, 2) residual connections, and 3) a feedforward network. }
    \label{fig:Transformer_intro}
\end{figure}

\begin{figure*}
    \centering
    \includegraphics[width=0.99\linewidth]{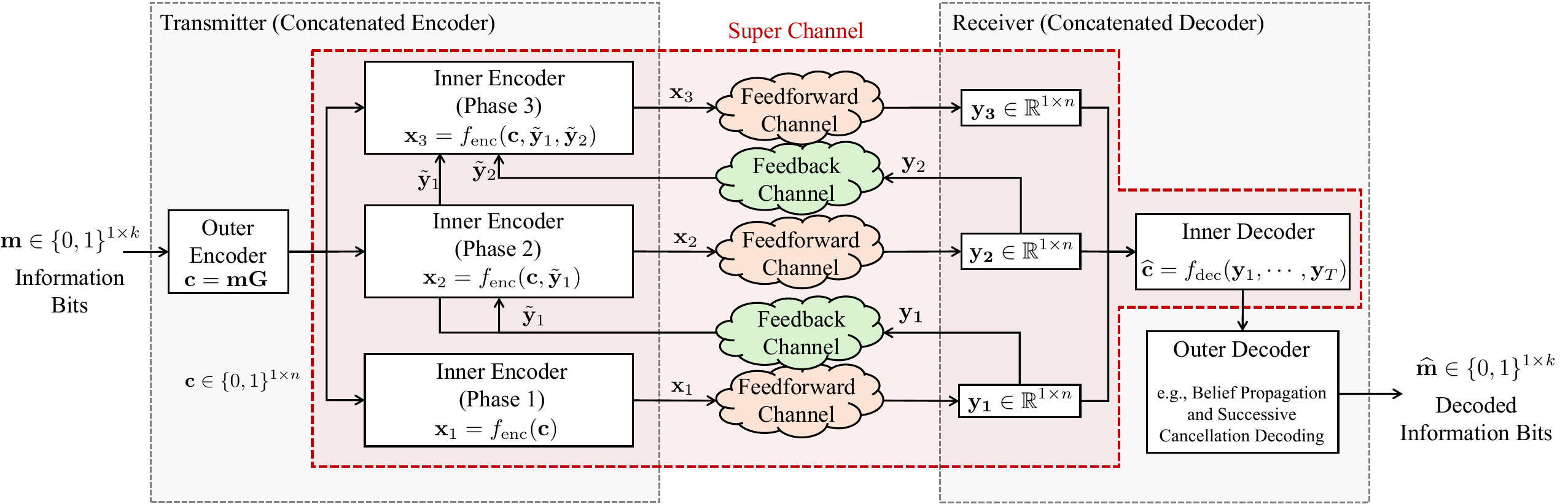}
    \caption{
        Illustration of the unfolded system model of the three-phase concatenated coding ($M=n\cdot T$, $T=3$). 
        Let us consider information bits with a length of $k$.
        First, the outer encoder generates an $n$-bit codeword $\mathbf{c}$ by \eqref{eq:cmG}.
        Second, the inner encoder exploits $T$ closed-loop iterations of transmitting parity messages $\mathbf{x}_t$ to the receiver.
        Third, the inner decoder works as a soft-decision estimator for the transmitted codeword $\mathbf{c}$, where the estimated codeword is $\widehat{\mathbf{c}}$. 
        Last, the outer decoder---a conventional linear block code decoder---forwards decoded information bits $\widehat{\mathbf{m}}$.
    }
    \label{fig:system_model}
\end{figure*}

\subsection{Transformer Layer} 

\MaskCode~builds upon the Transformer architecture~\cite{vaswani2017attention}, which we briefly introduce here.
The Transformer model consists of a couple of Transformer layers, which are depicted in \cref{fig:Transformer_intro}.
As shown in the right part of the figure, a Transformer layer consists of 1) a masked self-attention block, 2) residual connections and layer normalization (Add \& LayerNorm), and 3) a feedforward network (FFN).
The detailed operations of the residual connections and feedforward network are well known, so we focus on the masked self-attention block.

\paragraph*{Self-attention}
Let us consider an embedding $\mathbf{X}\in\mathbb{R}^{\ell \times d}$, where $\ell$ and $d$ denote the embedding dimension and the number of input tokens, respectively. 
The self-attention mechanism provides a way to compute a contextual representation of sequence $\mathbf{X}$ by relating input tokens in different positions~\cite{vaswani2017attention}.
It operates on three distinct matrices: the query $\mathbf{Q}$, key $\mathbf{K}$, and value $\mathbf{V}$.
These matrices are generated by applying learnable linear projections ($\mathbf{W}_Q$,$\mathbf{W}_K$, and $\mathbf{W}_V\in\mathbb{R}^{\ell \times \ell }$) to input $\mathbf{X}$:
\begin{equation}
    \mathbf{Q} = \mathbf{W}_Q\mathbf{X}, \quad \mathbf{K} = \mathbf{W}_K\mathbf{X}, \quad \mathbf{V} = \mathbf{W}_V\mathbf{X},
\end{equation}
where we follow a transposed convention from the original Transformer paper~\cite{vaswani2017attention} such that $\mathbf{Q}$, $\mathbf{K}$, and $\mathbf{V}$ are column-wise token representations.
The \textit{query} acts as a probe seeking relevant information, the \textit{key} serves as an index for matching, and the \textit{value} contains the actual feature content to be retrieved. 
Then, the self-attention output without masking can be obtained by $\mathtt{Attention}(\mathbf{Q}, \mathbf{K}, \mathbf{V}) = \mathbf{V}^\mathrm{T} \mathtt{softmax}\left(\frac{\mathbf{K}^\mathrm{T}\mathbf{Q}}{\sqrt{\ell}}\right)$, where the softmax function $\mathtt{softmax}(\cdot)$ is applied to each column of the matrix.

\paragraph*{Masked self-attention}
While the self-attention mechanism enables each token to aggregate information from all others, this fully connected interaction is not always desirable. 
For instance, in autoregressive language modeling, a token should not aggregate future tokens to maintain causality~\cite{vaswani2017attention}.
To control the information flow, a mask matrix ($\mathbf{M}\in\{-\infty,0\}^{d\times d}$) is generally used in a self-attention block, \ie,
\begin{equation}\label{eq:attention}
    \mathtt{Attention}(\mathbf{Q}, \mathbf{K}, \mathbf{V}) = \mathbf{V}^\mathrm{T}\mathtt{softmax}\left(\frac{\mathbf{K}^\mathrm{T}\mathbf{Q}}{\sqrt{\ell}} + \mathbf{M}\right).
\end{equation}
In \eqref{eq:attention}, setting $[\mathbf{M}]_{ij}=-\infty$ ensures that the $j$-th column of the output does not aggregate information from the $i$-th column of the value matrix $\mathbf{V}$.
By appropriately designing $\mathbf{M}$, task-specific structural prior knowledge, such as causality~\cite{vaswani2017attention} or code structure~\cite{choukroun2022error,park2024crossmpt}, can be directly embedded into the attention computation, allowing the Transformer layers to focus only on task-relevant interactions.

\section{System Model: Feedback-Assisted Concatenated Codes
\label{sec:system_model}}

In this section, we introduce the system model for \textit{feedback-assisted} concatenated coding, which integrates i) an \textit{outer ECC} for open-loop error control and ii) an \textit{inner feedback code} for utilizing channel output feedback.
The outer linear block code $\mathcal{C}\subseteq \{0,1\}^{1\times n}$ is characterized by a pair of matrices: a generator matrix $\mathbf{G} \in \{0,1\}^{k\times n}$ and a parity-check matrix $\mathbf{H} \in \{0,1\}^{(n-k) \times n}$, satisfying $\mathbf{G}\mathbf{H}^\mathrm{T}=\mathbf{0}_{k\times(n-k)}$ \cite{mceliece2002theory}.

\subsection{Channel Model}
We consider a feedback-enabled memoryless additive white Gaussian noise (AWGN) channel with $T$ closed-loop iterations, hereafter referred to as \textit{phases}.
The transmitter seeks to reliably send a $k$-bit message $\mathbf{m}$ by transmitting its encoded $n$-bit codeword $\mathbf{c}=\mathbf{mG}$ through the inner feedback code.
The feedforward and feedback channel noises are modeled as independent Gaussian random variables $\mathcal{N}(0,\sigma_\text{ff}^2)$ and $\mathcal{N}(0,\sigma_\text{fb}^2)$, where $\sigma_\text{ff}^2$ and $\sigma_\text{fb}^2$ denote the respective noise variances.
The inner code transmits an $n$-bit codeword $\mathbf{c}$ via $M$ channel uses, yielding an inner code rate of $\nicefrac{n}{M}$.
In the proposed method, we adopt the bit-based feedback protocol~\cite{sk1966coding-1,chance2011concatenated,shao2023attentioncode} over the sub-block-based feedback protocol~\cite{ozfatura2022all,ankireddy2025lightcode}; hence, $M=nT$.
At the $t$-th forward transmission phase, the transmitter sends an $n$-length vector $\mathbf{x}_t$, referred to as the parity message, subject to an average power constraint 
\begin{equation}
    \frac{1}{n}\mathbb{E}\left[\Vert\mathbf{x}_t\Vert^2\right] \le P,
\end{equation}
where $P$ denotes the average power constraint per phase.
Denoting the feedforward noise at the $t$-th phase as $\mathbf{n}_t\sim\mathcal{N}(0,\sigma_\text{ff}^2\mathbf{I}_n)$, the received signal $\mathbf{y}_t$ is given by 
\begin{equation}
    \mathbf{y}_t = \mathbf{x}_t + \mathbf{n}_t, ~\forall t\in[T].
\end{equation}

The receiver then sends $\mathbf{y}_t$ back to the transmitter via the feedback channel.
Following
~\cite{kim2018deepcode,pmlr-v202-kim23j,chance2011concatenated,sk1966coding-1,Yihan2020FeedbackTurbo}, we assume the \textit{passive feedback}, whereby the feedback signal received at the transmitter is given by 
\begin{equation}
    \tilde{\mathbf{y}}_t = \mathbf{y}_t + \mathbf{z}_t = \mathbf{x}_t+\mathbf{n}_t+\mathbf{z}_t, ~\forall t\in[T-1],
\end{equation}
where $\mathbf{z}_t\sim\mathcal{N}(0,\sigma_\text{fb}^2\mathbf{I}_n)$ denotes the feedback noise at the $t$-th phase.
With this channel model in place, we next define the encoding and decoding procedure of the inner feedback code.

\subsection{Inner Feedback Code}

The inner feedback code operates over the channel defined above, utilizing noisy feedback to iteratively refine the transmitted signal over $T$ phases.
Figure \ref{fig:system_model} depicts an unfolded diagram of the system model with $T=3$.
The outer encoder first generates an $n$-bit codeword $\mathbf{c}=\mathbf{mG}\in\mathcal{C}$, after which the inner encoder performs $T$ closed-loop transmission iterations. 
The inner decoder then provides the LLRs of the estimated codeword, which are subsequently passed to the outer decoder for BP decoding. 
We now formally define the encoding and decoding functions of the inner feedback code.

The inner encoder aims to generate parity message $\mathbf{x}_t$ from the codeword $\mathbf{c}$ and the previously received feedback parity messages $\tilde{\mathbf{y}}_1,\tilde{\mathbf{y}}_2,\ldots,\tilde{\mathbf{y}}_{t-1}.$ 
Here, we represent the inner encoding as a function $f_{\text{enc}}(\cdot)$, where the encoding process is given by 
\begin{equation}\label{eq:encode_x}
    \mathbf{x}_t = \begin{cases}
        f_{\text{enc}}(\mathbf{c}), & \text{if $t=1$}, \\
        f_{\text{enc}}(\mathbf{c},\tilde{\mathbf{y}}_1,\cdots, \tilde{\mathbf{y}}_{t-1}), & \text{otherwise}.
    \end{cases} 
\end{equation}

At the receiver, the decoder aims to get the estimated codeword $\widehat{\mathbf{c}}$ from the received parity messages $\mathbf{y}_1,\mathbf{y}_2,\ldots,\mathbf{y}_T$.
Similar to the inner encoder in \eqref{eq:encode_x}, we represent the inner decoder as a function $f_{\text{dec}}(\cdot)$.
Then, the inner decoding process is defined by 
\begin{equation}\label{eq:decode_c}
    \widehat{\mathbf{c}} = f_{\text{dec}}(\mathbf{y}_1,\cdots, \mathbf{y}_{T}).
\end{equation}

What remains is to specify how $f_\text{enc}$ and $f_\text{dec}$ are designed to ensure reliable transmission.
\begin{itemize}
    \item \textbf{Linear feedback codes:} Classical methods like SK~\cite{sk1966coding-1} and CL~\cite{chance2011concatenated} schemes define the encoder $f_{\text{enc}}(\cdot)$ as a linear combination of $\mathbf{c},\tilde{\mathbf{y}}_1,\tilde{\mathbf{y}}_2,\ldots,\tilde{\mathbf{y}}_{t-1}$. Similarly, the decoder function $f_{\text{dec}}(\cdot)$ is defined by a linear combination of $\mathbf{y}_1,\mathbf{y}_2,\ldots,\mathbf{y}_{T}$. 
    Despite their analytical tractability, they are strictly limited by their linear architecture.
    \item \textbf{ML-based feedback codes:} The ML-based methods leverage neural networks like RNNs or Transformers (\eg, DeepCode~\cite{kim2018deepcode}, AttentionCode~\cite{shao2023attentioncode}, RobustCode~\cite{pmlr-v202-kim23j}, and~\cite{ozfatura2022all}). These nonlinear functions can represent robust, high-dimensional signaling mappings that outperform the linear baselines.
\end{itemize}

\paragraph*{Compatibility with outer decoders}
For \textit{ML-based feedback codes}, the decoder DNNs are typically trained using the binary cross-entropy (BCE) loss. 
Notably, the decoder output $\widehat{\mathbf{c}}$ fed into the BCE loss is equivalent to a vector of LLRs of the transmitted bits.
Since the LLR output can be readily converted into hard decisions, the ML-based feedback codes are compatible with a wide range of outer decoders, including soft-decision decoders such as belief propagation~\cite{kschischang2001factor} and successive cancellation list decoding~\cite{tal2015list}, as well as hard-decision decoders like the Berlekamp-Massey algorithm~\cite{jl1967shift}.
To validate the effect of the end-to-end training across diverse inner feedback codes, we adopt a differentiable BP decoder as the outer decoder throughout this paper.

\begin{figure*}
    \centering
    \includegraphics[width=1\linewidth]{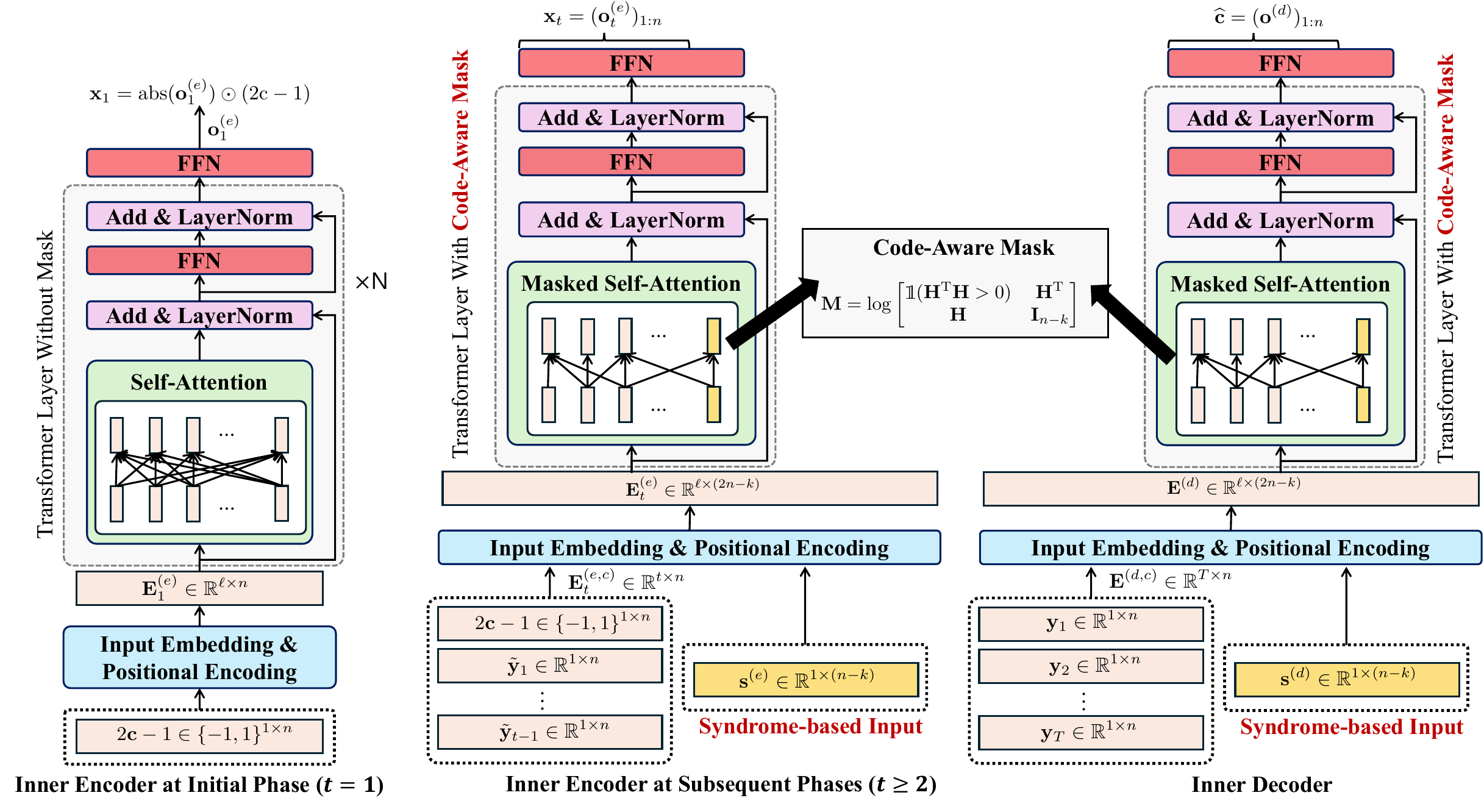}
    \caption{Overall schematic diagram of encoder/decoder of~\MaskCode. \textbf{(Left)} The inner encoder at initial phase ($t=1)$ processes the bipolar-mapped codeword using a standard Transformer. \textbf{(Middle)} The inner encoder at subsequent phases ($t\ge 2$) utilizes the \textit{code-aware mask} derived from the parity check matrix $\mathbf{H}$ and the \textit{soft syndrome-based input} to refine parity messages based on noisy feedback. \textbf{(Right)} Similar to the inner encoder, the inner decoder aims to produce reliable LLRs to the outer decoder by aggregating received parity messages, as well as the soft syndrome-based input corresponding to the initial-phase parity message. }
    \label{fig:maskcode_architecture}
\end{figure*}

\section{Proposed Method: \MaskCode}

In this section, we propose \MaskCode, a novel Transformer-based inner feedback code for the concatenated coding architecture in~\cref{sec:system_model}, where the schematic diagram is depicted in \cref{fig:maskcode_architecture}. 
The proposed method aims to address a fundamental limitation of existing approaches: the inner feedback code is typically designed independently of the outer code's structure, potentially misallocating feedback resources to error patterns already correctable by
the outer ECC. 
To this end, we incorporate the structural constraint of the outer code ($\mathbf{cH}^\mathrm{T}=\mathbf{0}_{1\times (n-k)}$) into the inner feedback code design via the following two key mechanisms: 
\begin{enumerate}
    \item \textbf{Soft syndrome-based input design} that informs the encoder about potential parity constraint violations.
    \item \textbf{Code-aware masking} strategy derived from the Tanner graph.
\end{enumerate}
In the remainder of this section, we introduce the Transformer-based encoding and decoding DNN designs corresponding to encoding~\eqref{eq:encode_x} and decoding~\eqref{eq:decode_c} functions in \cref{sec:system_model}.

\subsection{Inner Encoder: Initial Phase ($t=1$)}

As depicted in \cref{fig:maskcode_architecture}, at the initial phase, the inner encoder generates parity message $\mathbf{x}_1$ from codeword $\mathbf{c}$ produced by the outer encoder, as defined in \eqref{eq:encode_x}.

\paragraph*{Embedding}
To process $\mathbf{c}$ with the Transformer, we first map it into a high-dimensional token sequence. 
Specifically, the codeword is converted to the BPSK mapping, \ie, $(2\mathbf{c}-1)$ to ensure zero-mean symmetric inputs suitable for Transformer layers. 
The input embedding $\mathbf{E}_1^{(e)}\in\mathbb{R}^{\ell \times n}$ is then represented by
\begin{equation}\label{eq:embedding_p1}
    \mathbf{E}_1^{(e)} = \underbrace{(\mathbf{w}^{(e)}_1)^\mathrm{T}(2\mathbf{c}-1) }_{\text{Input Embedding}} + \underbrace{\mathbf{P}^{(e)}_1}_{\text{Positional Embedding}},
\end{equation}
where $\ell$ is the embedding dimension, and $\mathbf{w}^{(e)}_1\in\mathbb{R}^{1 \times \ell}$ and $\mathbf{P}^{(e)}_{1}\in\mathbb{R}^{\ell \times n}$ are learnable input embedding vector and positional embedding matrix, respectively.

\paragraph*{Parity message $\mathbf{x}_1$}
Since no feedback has been received at this phase, the encoder generates $\mathbf{x}_1$ without applying a code-aware mask. 
Let the output of the inner encoder's Transformer be $\mathbf{o}^{(e)}_1\in\mathbb{R}^{1\times n}$, \ie,
\begin{equation}
    \mathbf{o}^{(e)}_1 = \text{Trans}^{(e)}(\mathbf{E}^{(e)}_1; \mathbf{O}_n,\boldsymbol{\phi}^{(e)}_1),
\end{equation}
where $\mathbf{O}_n\in\mathbb{R}^{n\times n}$ and $\boldsymbol{\phi}^{(e)}_1$ denote the all-zero mask with size of $n$ and learnable parameters of the initial-phase encoder, respectively.
The parity message $\mathbf{x}_1$ is obtained by 
\begin{equation}\label{eq:init_phase_encode}
    \mathbf{x}_1 = \mathtt{abs}(\mathbf{o}_1^{(e)}) \odot (2\mathbf{c}-1),
\end{equation}
where $\odot$ denotes the Hadamard product. 
The \textit{key design rationale} behind \eqref{eq:init_phase_encode} is to explicitly preserve the sign of the BPSK-encoded codeword $(2\mathbf{c}-1)$ in the transmitted signal, thereby enabling the soft syndrome to be computed.
More importantly, this sign-fixing constraint is applied \textit{only at $t=1$}.
Extending this constraint to subsequent phases would force $\mathbf{x}_{t \geq 2}$ to share the sign of $(2\mathbf{c}-1)$, restricting the parity messages to amplitude-only modulation and degrading performance.
Hence, $\tilde{\mathbf{y}}_1$ serves as a fixed structural reference for the soft syndrome across all phases $t \geq 2$.


\subsection{Inner Encoder: Subsequent Phases $(t\ge 2)$}

Unlike the first phase, the encoder in subsequent phases can utilize the feedback history ($\tilde{\mathbf{y}}_1,\tilde{\mathbf{y}}_2,\ldots,\tilde{\mathbf{y}}_{t-1}$), which enables two core mechanisms of \MaskCode: i) a soft-syndrome-based input that identifies which parity constraints of the outer ECC are likely violated, and ii) a code-aware attention mask that restricts inter-bit interactions to follow the sparse structure of the Tanner graph. 
Together, these mechanisms allow the encoder to \textit{focus feedback 
resources on structurally important bits}, rather than treating all bits uniformly.

\paragraph*{Encoder input: code history}
As shown in \cref{fig:maskcode_architecture}, the first component of the encoder input aggregates the codeword $\mathbf{c}$ and all previously received feedback signals:
\begin{equation}\label{eq:input_conv}
    \mathbf{E}^{(e,c)}_t= 
    \begin{bmatrix}
        2\mathbf{c}-1 \\ 
        \tilde{\mathbf{y}}_1 \\
        \vdots \\
        \tilde{\mathbf{y}}_{t-1}
    \end{bmatrix}
    \in \mathbb{R}^{t \times n},
\end{equation}
This formulation follows the standard input design adopted in~\cite{kim2018deepcode,shao2023attentioncode,ozfatura2022all}. 

\paragraph*{Encoder input: soft syndrome}
The second component is the key novel input of \MaskCode. 
While existing feedback coding schemes rely solely on \eqref{eq:input_conv}, we introduce an additional input, the \textit{soft syndrome} $\mathbf{s}^{(e)}\in\mathbb{R}^{1\times (n-k)}$, which explicitly informs the encoder about which parity constraints of the outer ECC are likely being violated. 
To obtain $\mathbf{s}^{(e)}$, we first derive the LLRs of $\tilde{\mathbf{y}}_1$ by 
\begin{equation}\label{eq:LLR_encoder}
    \begin{split}
    \tilde{r}_{i} &= \log\left(
    \frac{\Pr(c_{i}=1|\tilde{y}_{1,i})}{\Pr(c_{i}=0|\tilde{y}_{1,i})}
    \right) \\ 
    &= \log \left(\frac{1+\mathrm{erf}\left(\frac{\tilde{y}_{1,i}}{\sqrt{2}\sigma_\text{eff}}\right)}{1-\mathrm{erf}\left(\frac{\tilde{y}_{1,i}}{\sqrt{2}\sigma_\text{eff}}\right)}\right),~\forall i\in[n],
\end{split}
\end{equation}
where $\sigma_{\text{eff}}^2=\sigma_{\text{ff}}^2+\sigma_{\text{fb}}^2$ and $\mathrm{erf}(x)=\frac{2}{\sqrt{\pi}}\int^{x}_0 \exp(-t^2)dt$.
Following the variable-to-check node update in the BP algorithm in \eqref{eq:vtoc}, the soft syndrome $\mathbf{s}^{(e)}\in\mathbb{R}^{1\times (n-k)}$ is then 
\begin{equation}\label{eq:syndrome_based_input_design}
    s^{(e)}_{i} = 2\mathrm{arctanh}\left(\prod_{b\in N_\text{v}(i)}\tanh\left(\frac{\tilde{r}_{b}}{2}\right)\right),~\forall i\in[n-k],
\end{equation}
where $N_\text{v}(i)=\{b\in[n]~|~[\mathbf{H}]_{i,b}=1\}$.
Here, $s^{(e)}_i\ll0$ indicates a probable violation of the $i$-th parity constraint, providing the encoder with structural information to focus feedback resources on the most vulnerable bits.

\paragraph*{Embedding}
By aggregating the code and syndrome parts of the encoder's input, we have the embedded input of the Transformer as
\begin{multline}\label{eq:input_design_enc}
    \mathbf{E}^{(e)}_t = 
    \left[\begin{array}{c|c}
        \mathbf{W}^{(e,c)}_t\mathbf{E}^{(e,c)}_t &
        (\mathbf{w}^{(e,s)}_t)^\mathrm{T}\mathbf{s}^{(e)}
    \end{array}\right] \\ + \mathbf{P}_{t}^{(e)}\in\mathbb{R}^{\ell \times (2n-k) },
\end{multline}
where $\mathbf{W}_t^{(e,c)}\in\mathbb{R}^{\ell\times t}$ and $\mathbf{w}_t^{(e,s)}\in\mathbb{R}^{1\times \ell}$ are learnable embedding matrix for code and learnable embedding vector for soft syndrome, respectively, and $\mathbf{P}_t^{(e)}\in\mathbb{R}^{\ell\times(2n-k)}$ denotes the positional embedding matrix.

\paragraph*{Code-aware attention mask}
The standard all-zero mask allows unrestricted interactions between all $n$ bits, which is agnostic to the code structure. 
\MaskCode~instead enforces a \textit{sparse, structure-aware} attention pattern derived from the Tanner graph, compelling the Transformer to respect the parity-check constraints $\mathbf{c}\mathbf{H}^\mathrm{T}=\boldsymbol{0}$ during encoding.
Tanner graph-based masking has been widely adopted in Transformer-based ECC decoders~\cite{choukroun2022error,levy2024accelerating,10960678,choukroun2024a}, demonstrating that aligning attention patterns with code structure significantly enhances decoding performance. 
However, to the best of our knowledge, this principle has never been applied to the \textit{encoding} side---specifically, to an inner feedback encoder that is explicitly aware of the outer ECC's structural constraints within the concatenated coding.

Following~\cite{choukroun2022error}, the code-aware attention mask $\mathbf{M} := g(\mathbf{H})$ is derived from the parity-check matrix $\mathbf{H}$, where $g:\{0,1\}^{(n-k)\times n}\rightarrow\{-\infty,0\}^{(2n-k)\times(2n-k)}$.
The code-aware attention mask enforces the Transformer to respect the code's structural constraints $\mathbf{c}\mathbf{H}^\mathrm{T}=\boldsymbol{0}$ during the attention computation.
The mask is constructed from a binary adjacency matrix $\mathbf{A} \in \{0,1\}^{(2n-k) \times (2n-k)}$, defined as the sum of two components.
\begin{itemize}
    \item \textbf{Inter-node connectivity ($\mathbf{A}_\text{inter}$):} The conventional bipartite adjacency matrix \cite{biggs1993algebraic, crnkovic2022ldpc} represents the explicit connections between variable nodes and check nodes by
        \begin{equation} \label{eq:inter-node connection matrix}
        \mathbf{A}_\text{inter} = 
        \left[
        \begin{array}{c|c}
        \mathbf{O}_{n} &  \mathbf{H}^\mathrm{T}\\ \hline
        \mathbf{H} & \mathbf{O}_{n-k}
        \end{array}
        \right].
        \end{equation}
    \item \textbf{Intra-node connectivity ($\mathbf{A}_\text{intra}$):} A block-diagonal matrix represents the induced connections within each node set by
        \begin{equation} \label{eq:intra-node connection matrix}
            \mathbf{A}_\text{intra} = 
            \left[
            \begin{array}{c|c}
            \mathbbm{1}(\mathbf{H}^\mathrm{T}\mathbf{H}>0) &  \mathbf{O}_{n \times (n-k)}\\ \hline
            \mathbf{O}_{(n-k) \times n} & \mathbf{I}_{n-k}
            \end{array}
            \right],
    \end{equation}
    where two variable nodes are connected if they share at least one common check node.
\end{itemize}
Summing these two components ($\mathbf{A}_\text{intra}$ and $\mathbf{A}_\text{inter}$) yields the final augmented adjacency matrix
\begin{equation}\label{eq:code-aware binary mask matrix}
\mathbf{A} = \mathbf{A}_\text{intra} + \mathbf{A}_\text{inter} 
=
\left[
\begin{array}{c|c}
\mathbbm{1}(\mathbf{H}^\mathrm{T}\mathbf{H}>0) &  \mathbf{H}^\mathrm{T}\\ \hline
\mathbf{H} & \mathbf{I}_{n-k}
\end{array}
\right],
\end{equation}
which integrates the Gram matrix-derived intra-node connectivity for variable nodes and the inter-node connectivity from the fundamental bipartite structure of ECC.
Finally, this binary matrix $\mathbf{A}$ is converted into the final attention mask $\mathbf{M}$ via an element-wise logarithm, which is then added to the attention scores just before the $\text{softmax}$ operation according to
\begin{equation} \label{eq:code-aware 0-inf mask matrix}
\mathbf{[M]}_{ij} := g(\mathbf{H})_{ij} = \log_2(\mathbf{A})_{ij}=
\begin{cases}
0,& \text{if }\mathbf{[A]}_{ij} =1,\\
-\infty, & \text{if }\mathbf{[A]}_{ij} =0.
\end{cases}
\end{equation}

\paragraph*{Parity message $\mathbf{x}_{t\ge2}$}
The Transformer output at the $t$-th phase is given by
\begin{equation}\label{eq:transformer_enc}
    \mathbf{o}_t^{(e)} = \text{Trans}^{(e)}(\mathbf{E}^{(e)}_t; \mathbf{M},\boldsymbol{\phi}^{(e)}_t),
\end{equation}
where $\boldsymbol{\phi}^{(e)}_t$ denotes the learnable parameters.
Since $\mathbf{o}_t^{(e)}$ contains representations for both variable nodes and check nodes, we extract only the first $n$ elements corresponding to the variable nodes:
\begin{equation}\label{eq:transformer_enc_output}
    \mathbf{x}_t = (\mathbf{o}_t^{(e)})_{1:n},
\end{equation}
where $(\mathbf{o}_t^{(e)})_{1:n}$ denotes extracting the first $n$ elements of $\mathbf{o}_t^{(e)}$.

\subsection{Inner Decoder}

The inner decoder mirrors the architecture of the subsequent encoder, employing the same code-aware attention mask $\mathbf{M}$ and soft syndrome-based input design.
After $T$ transmission phases, the receiver aggregates parity messages ($\mathbf{y}_1,\mathbf{y}_2,\ldots,\mathbf{y}_T$).
Then, the decoder leverages all the parity messages to estimate the original codeword $\mathbf{c}$. 

\paragraph*{Input embedding}

Similar to embedding in the subsequent encoder, the code and syndrome parts of the decoder's input are represented by 
\begin{itemize}
    \item \textbf{Code part}: Similar to \eqref{eq:input_conv}, the code part of the decoder's input is written as 
        \begin{equation}\label{eq:code_part_Dec}
            \mathbf{E}^{(d,c)}= 
            \begin{bmatrix}
                \mathbf{y}_1 \\ \vdots \\ \mathbf{y}_{T}
            \end{bmatrix}
            \in \mathbb{R}^{T \times n}.
        \end{equation}
    \item \textbf{Syndrome part}: The soft syndrome of the decoder is computed analogously to \eqref{eq:syndrome_based_input_design}, but using $\mathbf{y}_1$ with noise variance $\sigma_\text{ff}^2$:
        \begin{equation}
            \mathbf{s}^{(d)}\in\mathbb{R}^{1\times (n-k)},
        \end{equation}
        where $s^{(d)}_{i} = 2\mathrm{arctanh}\left(\prod_{b\in N_\text{v}(i)}\tanh\left(\frac{r_{b}}{2}\right)\right)$ and $r_{i} = \log \left(\frac{1+\mathrm{erf}\left(\frac{y_{1,i}}{\sqrt{2}\sigma_\text{ff}}\right)}{1-\mathrm{erf}\left(\frac{y_{1,i}}{\sqrt{2}\sigma_\text{ff}}\right)}\right)$. Note that $\sigma_\text{ff}^2$ is used here instead of $\sigma_\text{eff}^2$, since the decoder observes $\mathbf{y}_1$ directly without additional feedback noise.
\end{itemize}
Then, we have the embedded input for the decoder as
\begin{multline}\label{eq:input_design_dec}
    \mathbf{E}^{(d)} = \left[\begin{array}{c|c}
        \mathbf{W}^{(d,c)}\mathbf{E}^{(d,c)} & 
        (\mathbf{w}^{(d,s)})^\mathrm{T} \mathbf{s}^{(d)} 
    \end{array}\right] \\ + \mathbf{P}^{(d)}\in\mathbb{R}^{\ell \times (2n-k)},
\end{multline}
where $\mathbf{W}^{(d,c)}\in\mathbb{R}^{\ell\times T}$ and $\mathbf{w}^{(d,s)}\in\mathbb{R}^{1\times \ell}$ denote the learnable embedding matrix for the code and learnable embedding vector for soft syndrome, respectively, and $\mathbf{P}^{(d)}\in\mathbb{R}^{\ell\times(2n-k)}$ denotes the positional embedding matrix. 

\paragraph*{Decoding}

Let the output of the Transformer block be $\mathbf{o}^{(d)}\in\mathbb{R}^{1\times(2n-k)}$, \ie, 
\begin{equation}\label{eq:transformer_dec}
    \mathbf{o}^{(d)} = \text{Trans}^{(d)}(\mathbf{E}^{(d)}; \mathbf{M},\boldsymbol{\phi}^{(d)}),
\end{equation}
where $\boldsymbol{\phi}^{(d)}$ denotes the learnable parameters of the decoder.
Analogous to \eqref{eq:transformer_enc_output}, we extract the variable node outputs as the estimated codeword:
\begin{equation}
    \widehat{\mathbf{c}} = (\mathbf{o}^{(d)})_{1:n},
\end{equation}
where each element $\widehat{c}_i$ corresponds to the LLR of the 
transmitted bit $c_i$:
\begin{equation}
    \widehat{c}_i=\log\left(\frac{\Pr(c_i=1|\mathbf{y}_1,\mathbf{y}_2,\ldots,\mathbf{y}_T)}{\Pr(c_i=0|\mathbf{y}_1,\mathbf{y}_2,\ldots,\mathbf{y}_T)}\right).
\end{equation}

\subsection{Implementation and Loss Function}
\label{subsec:implementation_loss}

\MaskCode~can be trained end-to-end by minimizing a BCE loss that jointly considers the inner encoder, inner decoder, and the outer BP decoder.
Specifically, the inner decoder output $\widehat{\mathbf{c}}$ is passed through $N_\text{BP}$ BP iterations to yield $\widehat{\mathbf{c}}_\text{BP} = \mathtt{BP}(\widehat{\mathbf{c}};N_\text{BP})$, where $N_\text{BP}=0$ reduces to inner-code-only training.

For notational convenience, we aggregate the trainable parameters as follows:
\begin{itemize}
    \item $\boldsymbol{\theta}_1^{(e)}$: Trainable weights of the initial-phase encoder:
        \begin{equation}
            \boldsymbol{\theta}_1^{(e)} = \{\mathbf{w}_1^{(e)},\mathbf{P}_1^{(e)},\boldsymbol{\phi}_1^{(e)}\}.
        \end{equation}
    \item $\boldsymbol{\theta}_{t\ge2}^{(e)}$: Trainable weights of the subsequent-phase encoder:
        \begin{equation}
            \boldsymbol{\theta}_{t\ge2}^{(e)} = \{\mathbf{W}_t^{(e,c)},\mathbf{w}_t^{(e,s)},\mathbf{P}_t^{(e)},\boldsymbol{\phi}_t^{(e)}\}.
        \end{equation}
    \item $\boldsymbol{\theta}^{(d)}$: Trainable weights of the decoder:
        \begin{equation}
            \boldsymbol{\theta}^{(d)} = \{\mathbf{W}^{(d,c)},\mathbf{w}^{(d,s)},\mathbf{P}^{(d)},\boldsymbol{\phi}^{(d)}\}.
        \end{equation}
\end{itemize}

\begin{algorithm}[t]
	\caption{Training Algorithm of \MaskCode\label{alg:maskcode}}
    \Inp{Training parameters}{
        Total training step: $T_\text{train}$, batch size $B$, feedforward noise scale $\sigma_\text{ff}$, feedback noise scale $\sigma_\text{fb}$, and learning rate $\gamma$.
    }
    \Comment{Initialize neural network weights}
    \Init{Neural network weights}{
        Encoder weights: $\boldsymbol{\theta}^{(e)}_1,\boldsymbol{\theta}^{(e)}_2,\ldots,\boldsymbol{\theta}^{(e)}_{T}$.\;
        Decoder weights: $\boldsymbol{\theta}^{(d)}$.
        Mask $\mathbf{M}$ by Equation \eqref{eq:code-aware 0-inf mask matrix}
    }
    \For(){$\{1,...,T_\text{train}\}$}{
        \ParallelFor{$b$ in $\{1,\ldots,B\}$}{
            Draw $\mathbf{m}$ uniformly at random from $\{0,1\}^{1\times k}$.\;
            $\mathbf{c}\leftarrow\mathbf{m}\mathbf{G}$\;
            \Comment{Initial Phase}
                Get embedded input $\mathbf{E}_1^{(e)}$ by \eqref{eq:embedding_p1}\;
                $\mathbf{o}^{(e)}_1\leftarrow\text{Trans}^{(e)}(\mathbf{E}_1^{(e)};\mathbf{O}_n,\boldsymbol{\phi_1^{(e)}})$.\;
                $\mathbf{x}_1\leftarrow \mathtt{abs}(\mathbf{o}^{(e)}_1)\odot(2\mathbf{c}-1)$.\;
                $\mathbf{y}_{1}=\mathbf{x}_{1}+\mathbf{n}_1$, where $\mathbf{n}_1\sim\mathcal{N}(0,\sigma_\text{ff}^2\mathbf{I}_n)$\;
            \Comment{Subsequent Phases}
            \For(){$t$ in $\{2,\ldots,T\}$}{
                $\tilde{\mathbf{y}}_{t-1}=\mathbf{y}_{t-1}+\mathbf{z}_{t-1}$, where $\mathbf{z}_{t-1}\sim\mathcal{N}(0,\sigma_\text{fb}^2\mathbf{I}_n)$\;
                Get embedded input $\mathbf{E}_t^{(e)}$ by \eqref{eq:input_design_enc}.\;
                $\mathbf{o}_t^{(e)}\leftarrow\text{Trans}^{(e)}(\mathbf{E}_{t}^{(e)};\mathbf{M},\boldsymbol{\phi_{t}^{(e)}})$.\;
                $\mathbf{x}_t=(\mathbf{o}_t^{(e)})_{1:n}$.\;
                $\mathbf{y}_{t}=\mathbf{x}_{t}+\mathbf{n}_t$, where $\mathbf{n}_t\sim\mathcal{N}(0,\sigma_\text{ff}^2\mathbf{I}_n)$.
            }
            \Comment{Decoding}
            Get embedded input $\mathbf{E}^{(d)}$ by \eqref{eq:input_design_dec}.\;
            $\mathbf{o}^{(d)}\leftarrow\text{Trans}^{(d)}(\mathbf{E}^{(d)};\mathbf{M},\boldsymbol{\phi^{(d)}})$.\;
            $\widehat{\mathbf{c}}=(\mathbf{o}^{(d)})_{1:n}$.\;
            $\widehat{\mathbf{c}}_\text{BP} \leftarrow  \mathtt{BP}(\widehat{\mathbf{c}};N_\text{BP})$.\;
            $\mathrm{Loss}_b \leftarrow - \sum_{i\in[n]} \big(c_i\log(\mathtt{sig}(\widehat{c}_{\text{BP},i}))$ $+(1-c_i)\log(1-\mathtt{sig}(\widehat{c}_{\text{BP},i}))\big)$.

        }
        $\mathrm{Loss}(\boldsymbol{\theta}^{(e)}_1,\ldots,\boldsymbol{\theta}^{(e)}_T,\boldsymbol{\theta}^{(d)}) = \frac{1}{B}\sum_{b\in[B]}\mathrm{Loss}_b(\boldsymbol{\theta}^{(e)}_1,\ldots,\boldsymbol{\theta}^{(e)}_T,\boldsymbol{\theta}^{(d)})$\;
        \Comment{Update Parameters}
        $\boldsymbol{\theta}_t^{(e)}\leftarrow\boldsymbol{\theta}_t^{(e)}-\gamma\frac{\partial\mathrm{Loss}}{\partial\boldsymbol{\theta}_t^{(e)}}$ for all $t\in[T]$.\;$\boldsymbol{\theta}^{(d)}\leftarrow\boldsymbol{\theta}^{(d)}-\gamma\frac{\partial\mathrm{Loss}}{\partial\boldsymbol{\theta}^{(d)}}$.\;
    }
\end{algorithm}

The end-to-end loss function is defined as 
\begin{align}\label{eq:loss}
& \mathrm{Loss}(\boldsymbol{\theta}^{(e)}_1,\ldots,
\boldsymbol{\theta}^{(e)}_T,\boldsymbol{\theta}^{(d)})
\\
&= - \mathbb{E}\left[\sum_{i\in[n]} c_i\log(\mathtt{sig}(\widehat{c}_{\text{BP},i}))
+(1-c_i)\log(1-\mathtt{sig}(\widehat{c}_{\text{BP},i}))\right],\nonumber
\end{align}
where $\mathtt{sig}(\cdot)$ denotes the sigmoid function and $\widehat{c}_{\text{BP},i}$ is the $i$-th element of $\widehat{\mathbf{c}}_{\text{BP}}$.
The full training procedure is summarized in 
Algorithm~\ref{alg:maskcode}.

While $N_\text{BP}>0$ allows the inner code to account for the BP decoder's behavior, we empirically observe that $N_\text{BP}=0$ achieves the best performance (see \cref{fig:end-to-end-training}). 
We attribute this to the fact that \MaskCode~already incorporates, via the code-aware attention mask and soft syndrome-based input, the structural constraint of the outer ECC that end-to-end training attempts to instill.
Consequently, the gradient instability introduced by unrolling BP iterations only degrades training without offering additional gain.

\begin{figure}
    \centering
    \includegraphics[width=0.9\linewidth]{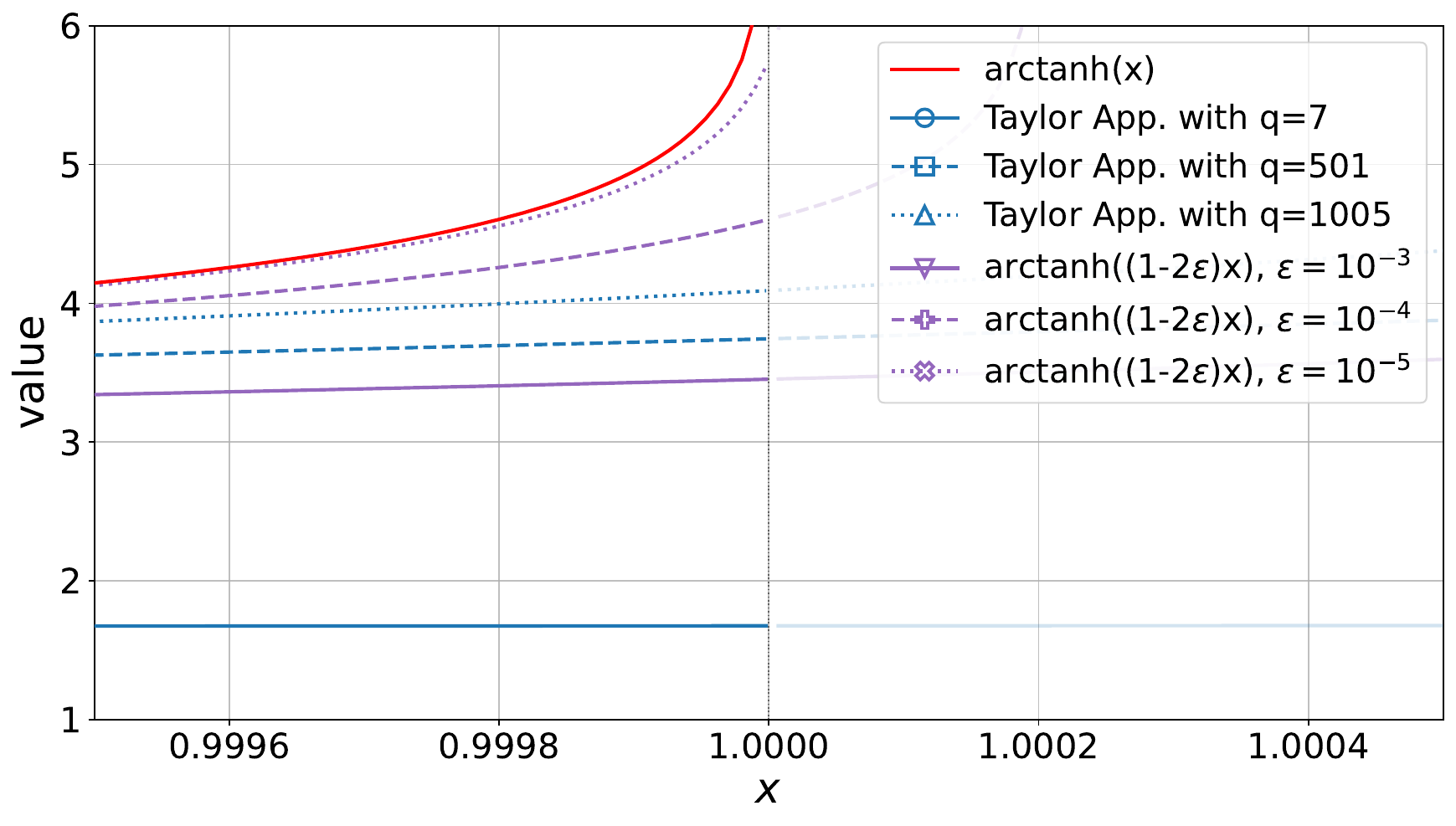}
    \caption{Stabilized $\mathrm{arctanh}$ function and Taylor approximation of the $\mathrm{arctanh}$ function. The order of Taylor approximation is denoted by $q$.}
    \label{fig:atanh_approx}
\end{figure}

\begin{remark}[Gradient instability in end-to-end BP training]
\label{rem:gradient_explode}
Backpropagation through unrolled BP iterations involves repeated differentiation of the $\tanh$ and $\mathrm{arctanh}$ operations.
Denoting the output of the $b$-th BP iteration as $g^{(b)}$ 
with $g^{(0)} = \widehat{\mathbf{c}}$, the gradient of the loss with 
respect to $\widehat{\mathbf{c}}$ is given by the chain rule:
\begin{equation}
    \frac{\partial \widehat{\mathbf{c}}_\text{BP}}{\partial \widehat{\mathbf{c}}} 
    = \mathbf{J}^{(N_\text{BP})} \mathbf{J}^{(N_\text{BP}-1)} 
    \cdots \mathbf{J}^{(1)},
\end{equation}
where $\mathbf{J}^{(b)} = \frac{\partial g^{(b)}}{\partial g^{(b-1)}}$ 
denotes the Jacobian of the $b$-th BP iteration.
As BP iterations progress, the magnitude of LLRs tends to grow, driving the inputs to $\tanh$ toward saturation ($|\mu_\text{vc}^{k\rightarrow j}| \to \infty$) and consequently pushing the inputs to $\mathrm{arctanh}$ toward $\pm 1$. 
In this regime, $\frac{\partial}{\partial x}\mathrm{arctanh}(x) = \frac{1}{1-x^2} \to \infty$ as $x \to \pm 1$, causing the Jacobians $\mathbf{J}^{(b)}$ to explode, as also mentioned in \cite{larue2021blind,nachmani2019hyper}. 
As this effect compounds across $N_\text{BP}$ iterations, $\mathbf{J}^{(N_\text{BP})}\cdots\mathbf{J}^{(1)}$ becomes increasingly ill-conditioned, making stable gradient-based optimization infeasible for large $N_\text{BP}$.
To address this instability, prior works~\cite{larue2021blind,nachmani2019hyper} approximate $\mathrm{arctanh}$ with a high-order Taylor polynomial, where the order needs to be higher than 1,000 to obtain a reasonable approximation, incurring substantial computational overhead per iteration.
Instead, we adopt a simple stabilizing strategy, $\mathrm{arctanh}((1-2\epsilon)x)$, which bounds the output without requiring any iteration.
As shown in \cref{fig:atanh_approx}, the stabilized function closely matches the exact $\mathrm{arctanh}$ over $|x| < 1$.
Combined with this stabilization, the end-to-end training of DeepCode~\cite{kim2018deepcode}, GBAF~\cite{ozfatura2022all}, and AttentionCode~\cite{shao2023attentioncode} shows the best performance with $N_\text{BP} = 2$ or $4$ iterations, which is comparable to the typical setting of $N_\text{BP} = 5$ used in~\cite{larue2021blind,nachmani2019hyper}, as will be shown in \cref{fig:end-to-end-training}. In our numerical results, we use $\epsilon=10^{-7}$, chosen via grid search over $\epsilon\in\{10^{-3},10^{-4},10^{-5},10^{-6},10^{-7},10^{-8},10^{-9}\}$.
\end{remark}

\section{Experimental Results}

In this section, we numerically evaluate \MaskCode~in a feedback-enabled communication scenario, by varying the additive noise power on the feedforward and feedback channels.
We begin by analyzing the effect of the end-to-end training with differentiable BP iterations (\cref{subsec:implementation_loss}).
We then compare \MaskCode~with existing feedback coding baselines for two outer codes: 1) BCH(31,16) and 2) LDPC(49,24), which follow the implementation of previous studies~\cite{park2024crossmpt,choukroun2022error}.
Subsequently, an ablation study validates the contribution of each core feature. 
Finally, we evaluate robustness under various feedback noise levels and compare computational complexity.

\subsection{Experimental Setup}

\begin{figure}
    \centering
    \subfloat[BCH(31,16) Code.\label{subfig:mask_bch_31_16}]{\includegraphics[width=0.5\linewidth]{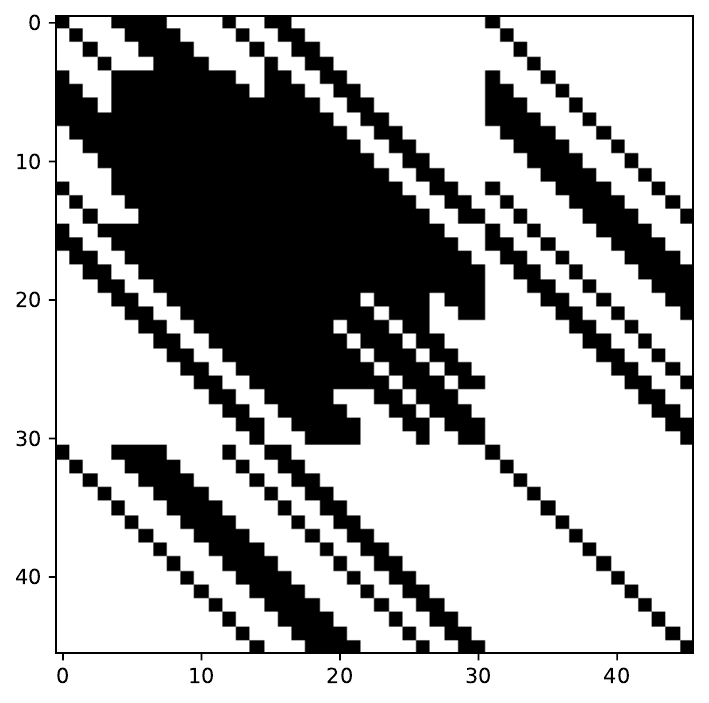}}
    \subfloat[LDPC(49,24) Code.\label{subfig:mask_ldpc_49_24}]{\includegraphics[width=0.5\linewidth]{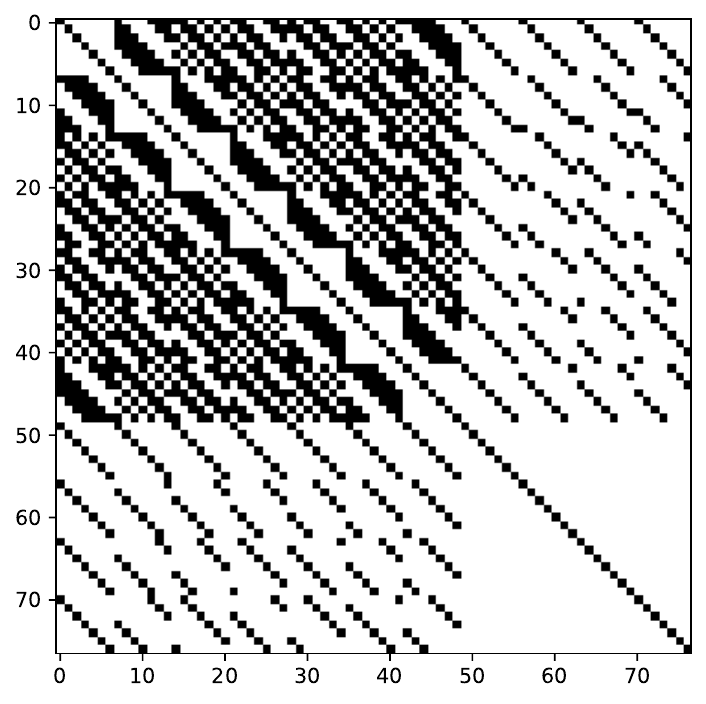}}
    \caption{Illustrations of the mask design by \eqref{eq:code-aware 0-inf mask matrix} for \protect\subref*{subfig:mask_bch_31_16}: BCH(31,16) outer code and \protect\subref*{subfig:mask_ldpc_49_24}: LDPC(49,24) outer code. 
    \label{fig:mask_design_results}}
\end{figure}

For the experiments, we consider a feedback-enabled memoryless AWGN channel, following the standard setup adopted in~\cite{kim2018deepcode,pmlr-v202-kim23j,shao2023attentioncode,ozfatura2022all,ankireddy2025lightcode,choukroun2022error}.
For the outer codes, we use the BP decoder (in \cref{subsec:ECCs}) as the outer decoder.
The attention masks for the outer codes are illustrated in \cref{fig:mask_design_results}.

\paragraph*{Setup for channels} 
The number of closed-loop iterations is three (\ie, $T=3$). 
We note that the power is normalized to $P=1$.
The feedforward and feedback channel SNRs are defined as $1/\sigma_\text{ff}^2$ and $1/\sigma_\text{fb}^2$, respectively, following the implementations of previous studies~\cite{kim2018deepcode,ankireddy2025lightcode,shao2023attentioncode,ozfatura2022all}.
Without any specification, the feedback SNR is assumed to be 20 dB, \ie, $\sigma^2_{\text{fb}}=0.01$.
The feedforward SNR varies from -5 dB to -1 dB for the BCH(31,16) outer code and from -5 dB to -2 dB for the LDPC(49,24) code.

\paragraph*{Training of \MaskCode} 

We use the \texttt{AdamW} optimizer, which extends the \texttt{Adam} optimizer with decoupled weight decay for improved generalization.
For the learning rate scheduling, we use \texttt{CosineAnnealingLR}, where the learning rate gradually decreases from $10^{-3}$ to $3\times 10^{-7}$.
The model is trained for 120,101 gradient steps, where the batch size is 2,048.
Unless otherwise specified, \MaskCode~uses two Transformer layers.

\paragraph*{Baselines}

For comparison, we will consider the following baseline schemes:
\begin{itemize}
    \item \textbf{Repetition code:} A baseline method where the transmitter repeatedly sends the $2\mathbf{c}-1$ for $T$ phases, which is denoted by Repeat.
    \item \textbf{SK scheme~\cite{sk1966coding-1,sk1966coding-2}:} SK scheme, denoted by SKCode, is a classical linear feedback coding scheme, which assumes noiseless feedback channels.
    \item \textbf{CL scheme~\cite{chance2011concatenated}:} CL scheme, denoted by CLCode, is an optimal linear feedback scheme for noisy feedback channels.
    \item \textbf{Deepcode~\cite{kim2018deepcode}:} DeepCode is a deep learning-based feedback coding scheme, which models the transmitter and receiver as RNNs. 
    \item \textbf{GBAF~\cite{ozfatura2022all}:} GBAF scheme is a sub-block-based feedback coding scheme, where it uses an attention mechanism to generate parity messages. 
    \item \textbf{Attentioncode~\cite{shao2023attentioncode}:} AttentionCode is a bit-based feedback coding scheme that employs a Transformer architecture.
    \item \textbf{Robustcode~\cite{pmlr-v202-kim23j}:} RobustCode is an RNN-based scheme.
    This scheme utilizes $nT$ phases, because a single scalar value is transmitted per phase.
    \item \textbf{Syndromecode~\cite{11443900}:} SyndromeCode is our previous work, a Transformer-based feedback coding scheme that incorporates a soft syndrome-based input design.
    Unlike \MaskCode, it does not employ a code-aware attention mask; hence, it corresponds to the ablation configuration with soft syndrome input but without masking (\cref{tab:ablation_bch_31_16,tab:ablation_ldpc_49_24}).
\end{itemize}
We note that all baselines are trained from scratch for each feedforward SNR setting, and neither SNR mismatch nor curriculum learning~\cite{ozfatura2022all,shao2023attentioncode} is adopted for fair comparison.

\subsection{End-to-End Training with Differentiable Outer Decoder}

\begin{figure}[t]
    \centering
    \includegraphics[width=0.9\linewidth]{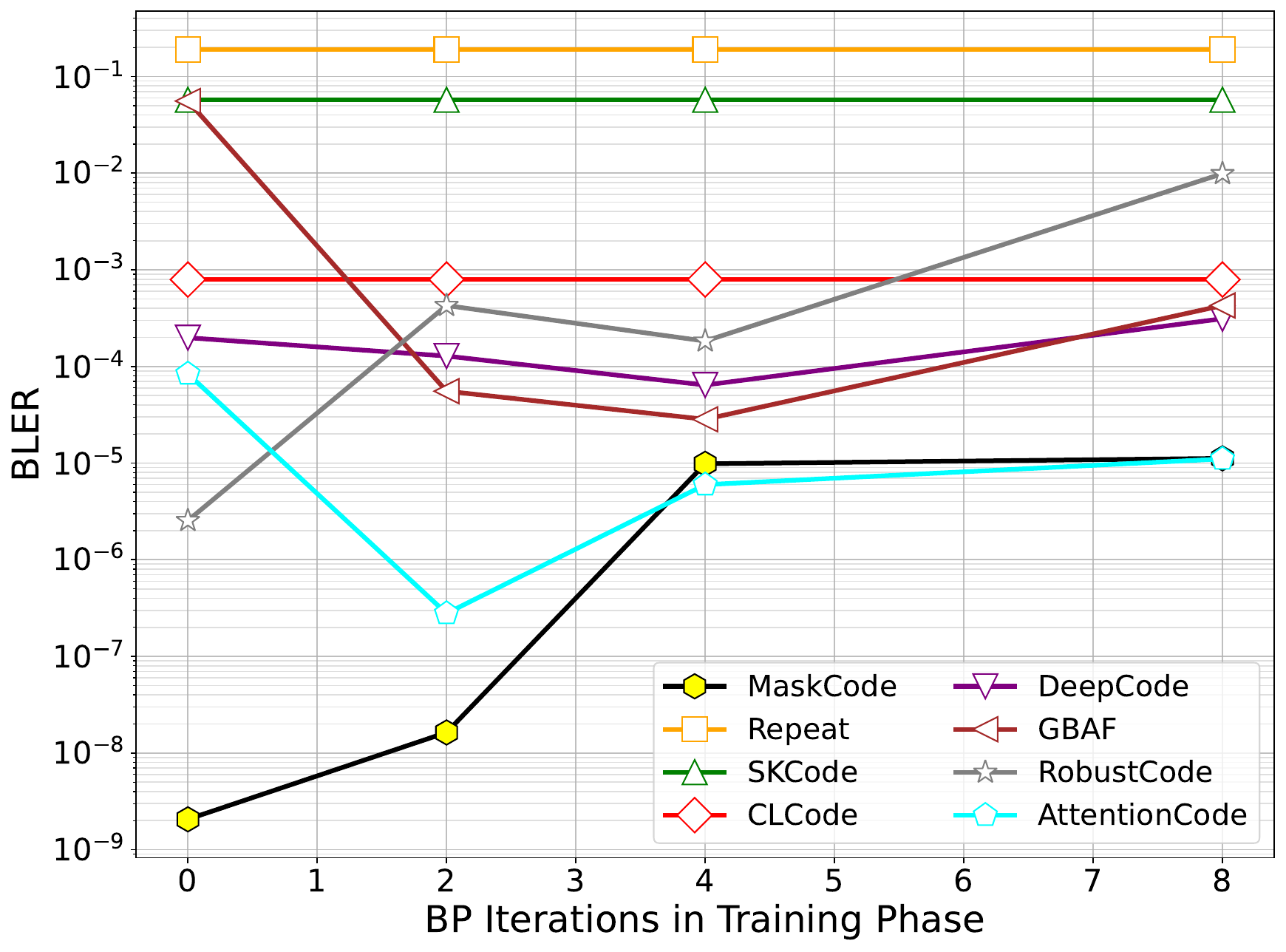}
    \caption{BLER evaluation of \MaskCode~and baselines under various numbers of BP iterations in the training phase. 
    The forward and feedback SNRs are assumed to be -2 dB and 20 dB, respectively. The outer code is the BCH(31,16) code.}
    \label{fig:end-to-end-training}
\end{figure}

As we discussed in \cref{subsec:implementation_loss}, \MaskCode~supports end-to-end training by incorporating the differentiable BP decoder into the training loop, where the number of BP iterations is $N_\text{BP}$. 
Although we stabilize the gradient computation in the BP decoder by stabilizing the $\mathrm{arctanh}$ function, the backpropagation through BP iterations still causes gradient explosion (see  \cref{rem:gradient_explode}), disrupting training convergence. 
To ensure stable training, we discard gradient updates whenever gradient explosion is detected, \ie, when the gradient norm is infinite.
Figure~\ref{fig:end-to-end-training} illustrates the BLER of various models with different $N_\text{BP}$, with the test phase fixed at 20 iterations for all cases.
Although previous studies typically use 50 BP iterations for evaluation~\cite{choukroun2022error,park2024crossmpt}, 20 BP iterations are sufficient in our experiments, as shown in \cref{fig:iter_BP}.

For the baseline ML-based schemes such as AttentionCode~\cite{shao2023attentioncode}, GBAF~\cite{ozfatura2022all}, and DeepCode~\cite{kim2018deepcode}, incorporating a small number of BP iterations (\eg, 2 or 4) during training yields a noticeable BLER improvement over the $N_\text{BP}=0$ case, suggesting that end-to-end training helps structurally agnostic models adapt to the outer code.
However, as $N_\text{BP}$ increases, performance degrades significantly due to gradient instability. 

In contrast, \MaskCode~achieves superior performance at $N_\text{BP}=0$, without any end-to-end BP training. 
Unlike the baselines that rely on the training process to internalize the outer code's structure, \MaskCode~explicitly incorporates the outer ECC's structure via its \textit{soft syndrome-based input} and \textit{code-aware attention mask}---the structural constraint of the outer code that end-to-end training aims to incorporate. 
Consequently, adding BP iterations to \MaskCode's training only introduces gradient instability, leading to performance degradation.
These results demonstrate that an outer code-aware architecture is more robust and effective than relying on end-to-end training.
Hence, in the remainder of this paper, we report results with $N_\text{BP}=0$. 

\begin{figure}
    \centering
    \subfloat[BCH(31,16) Outer Code. The effective code rate is $\nicefrac{n}{kT}\approx 0.17$\label{subfig:main_bch_31_16}]{\includegraphics[width=0.9\linewidth]{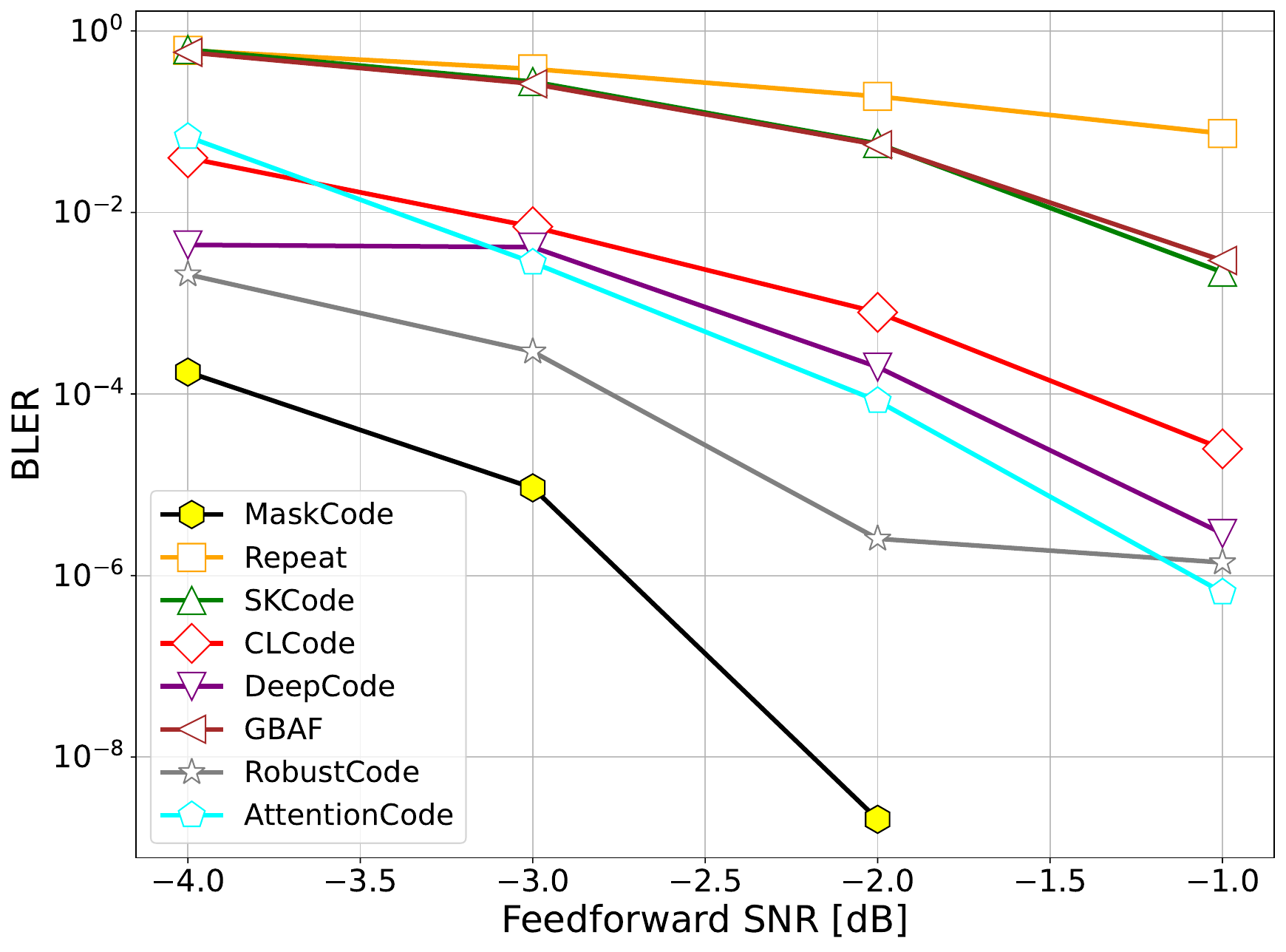}}
    \\
    \subfloat[LDPC(49,24) Outer Code. The effective code rate is $\nicefrac{n}{kT}\approx 0.16$\label{subfig:main_ldpc_49_24}]{\includegraphics[width=0.9\linewidth]{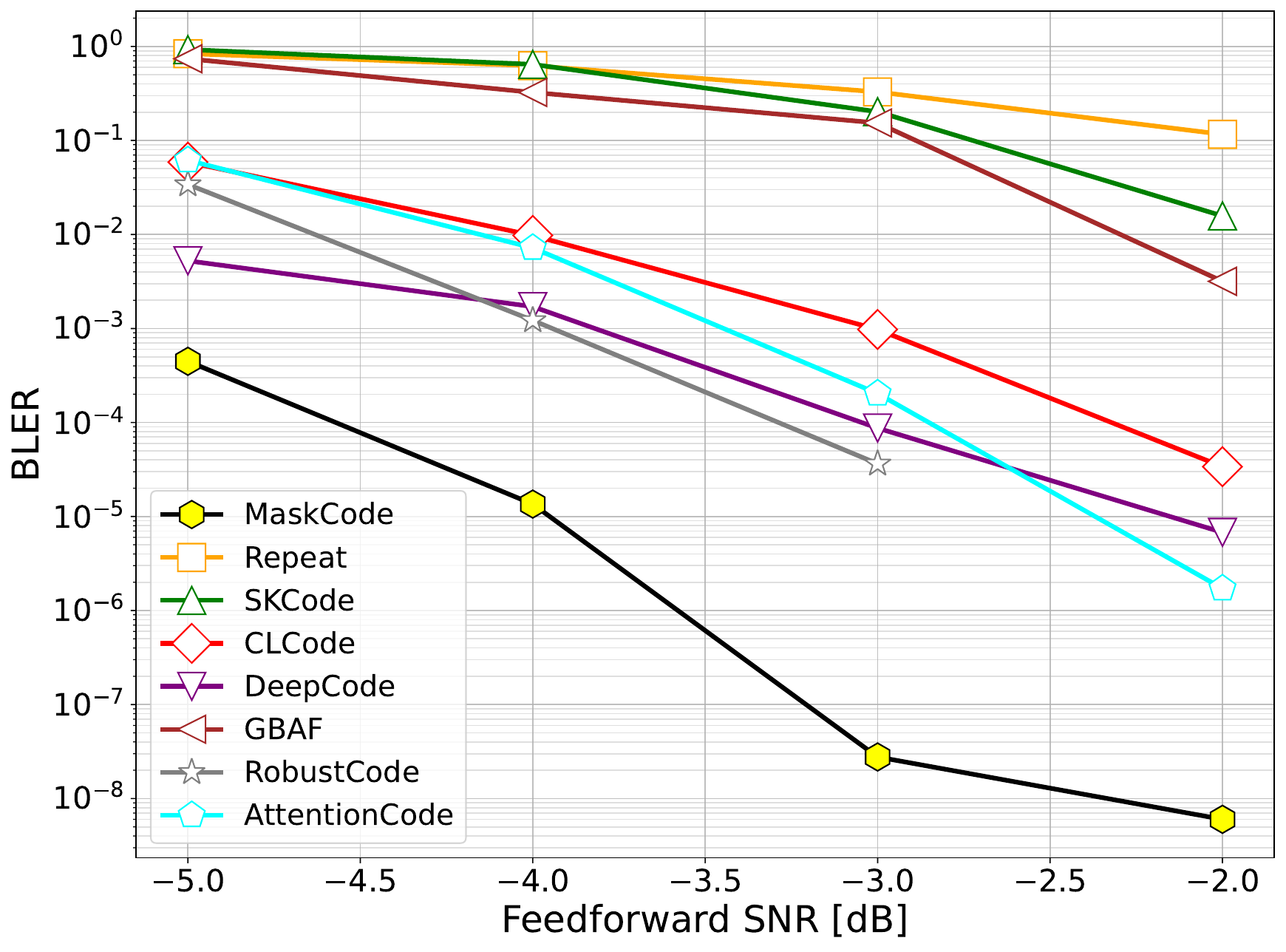}}
    \caption{BLER performance of \MaskCode~and baseline schemes for a 3-phase feedback-assisted concatenated code\label{fig:main}}
\end{figure}

\subsection{BLER Performance under Varying Feedforward SNR}

In this subsection, we evaluate \MaskCode~by varying the feedforward noise power $\sigma_\text{ff}^2$ for two outer codes: BCH(31,16) and LDPC(49,24), with results shown in \cref{fig:main}.
As shown in \cref{subfig:main_bch_31_16}, \MaskCode~outperforms all baselines across the entire SNR range, achieving an SNR gain of approximately 1.0 dB over the best-performing baseline at the same target BLER for BCH(31,16).
In \cref{subfig:main_ldpc_49_24}, \MaskCode~consistently outperforms the baselines with an SNR gain of approximately 1.5 dB. 
Notably, this large gain---compared to the 1.0 dB gain observed for BCH(31,16)---is achieved despite the two outer codes having similar effective code rates, suggesting that the gain is driven by the structural properties of the Tanner graph rather than by code rate differences.
A sparser Tanner graph means each check node involves fewer variable nodes, so the code-aware mask enforces a more restrictive attention pattern compared to the fully-connected baseline---diverging more significantly from model-agnostic approaches that treat all bits uniformly, and thus yielding a larger performance gain.


Another notable observation is that GBAF~\cite{ozfatura2022all}, a sub-block-based method, exhibits substantial BLER degradation compared to bit-based methods.
This is because sub-block-based methods reinforce dependencies among adjacent bits within each sub-block, rather than across the entire codeword, causing the per-bit LLRs provided to the outer BP decoder to deviate from their true values.
Consequently, the outer decoder receives structurally inconsistent LLRs, leading to significant performance degradation in the concatenated coding scenario.

\subsection{Ablation Study: Code-Aware Mask \& Soft Syndrome Input}
\label{subsec:ablation_study}

\begin{table}[t]
    \centering
    \caption{Ablation study of the proposed method. The outer code is BCH(31,16) }
    \label{tab:ablation_bch_31_16}
    \adjustbox{width=1\linewidth}{
    \begin{tabular}{cccccccc}
    \toprule
        
        \multirow{1}{*}{Methods} & \multicolumn{2}{c}{Features} & \multicolumn{3}{c}{BLER}
        \\ \cmidrule(lr){2-3}\cmidrule(lr){4-6}
        & \makecell{Mask} & \makecell{Soft syndrome \\ Input} & \makecell{$@$SNR=-2dB} & \makecell{$@$SNR=-3dB} & \makecell{$@$SNR=-4dB} \\ 
        \midrule
        \multirow{4}{*}{\makecell{\MaskCode \\ \textbf{(Ours)}}} & \checkmark & \checkmark & \textbf{\underline{2.06$\times 10^{-9}$}} & \textbf{\underline{9.25$\times 10^{-6}$}} & \textbf{\underline{1.74$\times 10^{-4}$}} \\ 
         & \checkmark & \xmarkg &  \textbf{2.16$\times 10^{-6}$} & 3.97$\times 10^{-4}$ & 8.24$\times 10^{-3}$ \\ 
         & \xmarkg & \checkmark & 8.04$\times 10^{-6}$ & \textbf{2.64$\times 10^{-4}$} & \textbf{3.51$\times 10^{-4}$}\\ 
         & \xmarkg & \xmarkg & 1.20$\times10^{-5}$ & 1.28$\times10^{-3}$ & 1.58$\times10^{-2}$ \\
        \midrule
        DeepCode~\cite{kim2018deepcode} & - & - & 1.98$\times10^{-4}$ & 4.15$\times10^{-3}$ & 4.39$\times10^{-3}$\\
        RobustCode~\cite{pmlr-v202-kim23j} & - & - & 2.54$\times10^{-6}$ & 2.93$\times10^{-4}$ & 2.08$\times10^{-3}$ \\
        AttentionCode~\cite{ozfatura2022all} & - & - & 8.39$\times10^{-5}$ & 2.81$\times10^{-3}$ & 6.81$\times10^{-2}$ \\
    \bottomrule 
    \multicolumn{6}{l}{*The best method is marked in \textbf{\underline{bold and 
    underline}}.}\\
    \multicolumn{6}{l}{**The second-best method is marked in \textbf{bold}. }
    \end{tabular}
    }
\end{table}

\begin{table}
    \centering
    \caption{Ablation study of the proposed method. The outer code is LDPC(49,24) }
    \label{tab:ablation_ldpc_49_24}
    \adjustbox{width=1\linewidth}{
    \begin{tabular}{cccccccc}
    \toprule
        
        \multirow{1}{*}{Methods} & \multicolumn{2}{c}{Features} & \multicolumn{3}{c}{BLER}
        \\ \cmidrule(lr){2-3}\cmidrule(lr){4-6}
        & \makecell{Mask} & \makecell{Soft syndrome \\ Input} & \makecell{$@$SNR=-3dB} & \makecell{$@$SNR=-4dB} & \makecell{$@$SNR=-5dB} \\ 
        \midrule
        \multirow{4}{*}{\makecell{\MaskCode \\ \textbf{(Ours)}}} & \checkmark & \checkmark & \textbf{\underline{2.77$\times 10^{-8}$}} & \textbf{\underline{1.36$\times 10^{-5}$}} & \textbf{\underline{4.48$\times 10^{-4}$}} \\ 
         & \checkmark & \xmarkg & \textbf{1.53$\times 10^{-5}$} & 6.41$\times 10^{-4}$ & \textbf{4.39$\times 10^{-3}$} \\ 
         & \xmarkg & \checkmark & 2.22$\times 10^{-5}$ & 6.10$\times 10^{-4}$ & 6.47$\times 10^{-3}$\\ 
         & \xmarkg & \xmarkg & 2.81$\times 10^{-5}$ & \textbf{5.09$\times 10^{-4}$} & 8.42$\times 10^{-3}$ \\
        \midrule
        DeepCode~\cite{kim2018deepcode} & - & - & 1.60$\times 10^{-4}$ & 1.89$\times 10^{-3}$ & 5.74$\times 10^{-3}$ \\
        RobustCode~\cite{pmlr-v202-kim23j} & - & - & 3.64$\times 10^{-5}$ & 1.22$\times 10^{-3}$ & 3.41$\times 10^{-2}$ \\
        AttentionCode~\cite{ozfatura2022all} & - & - & 2.03$\times 10^{-4}$ & 7.20$\times 10^{-3}$ & 6.15$\times 10^{-2}$ \\
    \bottomrule
    \multicolumn{6}{l}{*The best method is marked in \textbf{\underline{bold and 
    underline}}.}\\
    \multicolumn{6}{l}{**The second-best method is marked in \textbf{bold}. }
    \end{tabular}
    }
\end{table}

Tables \ref{tab:ablation_bch_31_16} and \ref{tab:ablation_ldpc_49_24} detail the results of our ablation study, which is designed to deconstruct the performance of our proposed \MaskCode~and quantify the \textit{individual contribution} of its two key features: 1) the code-aware attention mask and 2) the soft syndrome-based input design. 
In this study, we evaluate and compare the performance of the following four model configurations: 1) conventional Transformer without either features, 2) soft syndrome-based input only, 3) code-aware mask only, and 4) \MaskCode. 
As shown in the tables, each feature independently yields a substantial improvement over the baseline, and their combination in the full \MaskCode~ model achieves the best performance, confirming that the two features are complementary and produce a synergistic effect.

\subsection{Various Numbers of Outer Code Iterations}

\begin{figure}[t]
    \centering
    \includegraphics[width=0.9\linewidth]{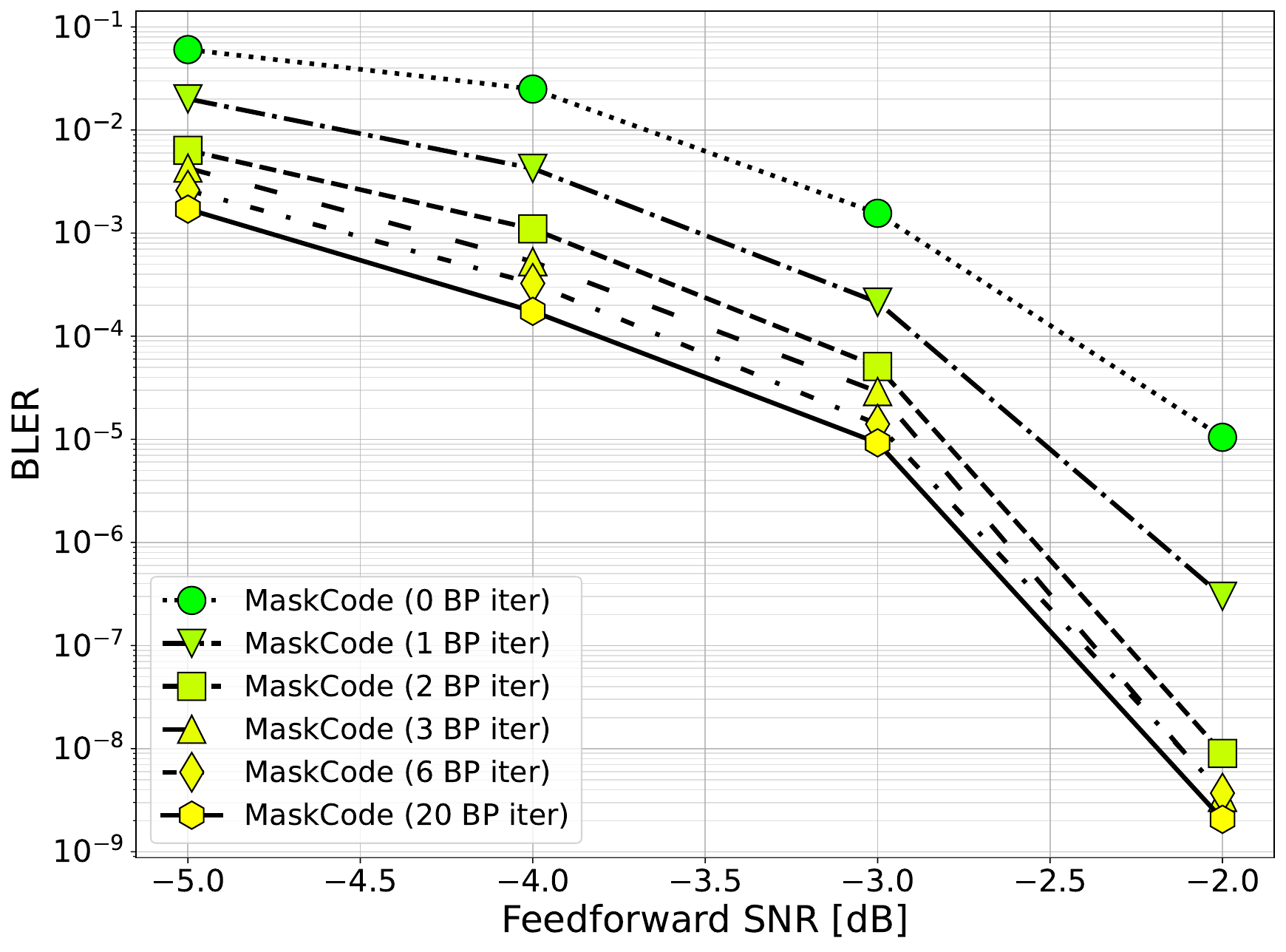}
    \caption{BLER performance of \MaskCode~for various numbers of outer decoder (BP) iterations. The \MaskCode~with few iterations (2 or 3) achieves a comparable BLER performance to that with 20 iterations. The outer code is the BCH(31,16) code.}
    \label{fig:iter_BP}
\end{figure}

To determine how many BP iterations are necessary for the inference, we evaluate the BLER across different numbers of BP iterations ($0$ to $20$) in \cref{fig:iter_BP}.
Without any BP iterations, the output relies solely on the inner decoder's LLRs, yielding a relatively high BLER.
A significant improvement is observed at $1$ and $2$ iterations, after which the performance gap becomes negligible up to $20$ iterations.
This indicates that only a few BP iterations are sufficient for near-optimal decoding performance in the concatenated system.

\subsection{Various Feedback Noise Scales}

\begin{figure}
    \centering
    \subfloat[BCH(31,16) Outer Code.\label{subfig:fb_snr_sweep_bch}]{\includegraphics[width=0.9\linewidth]{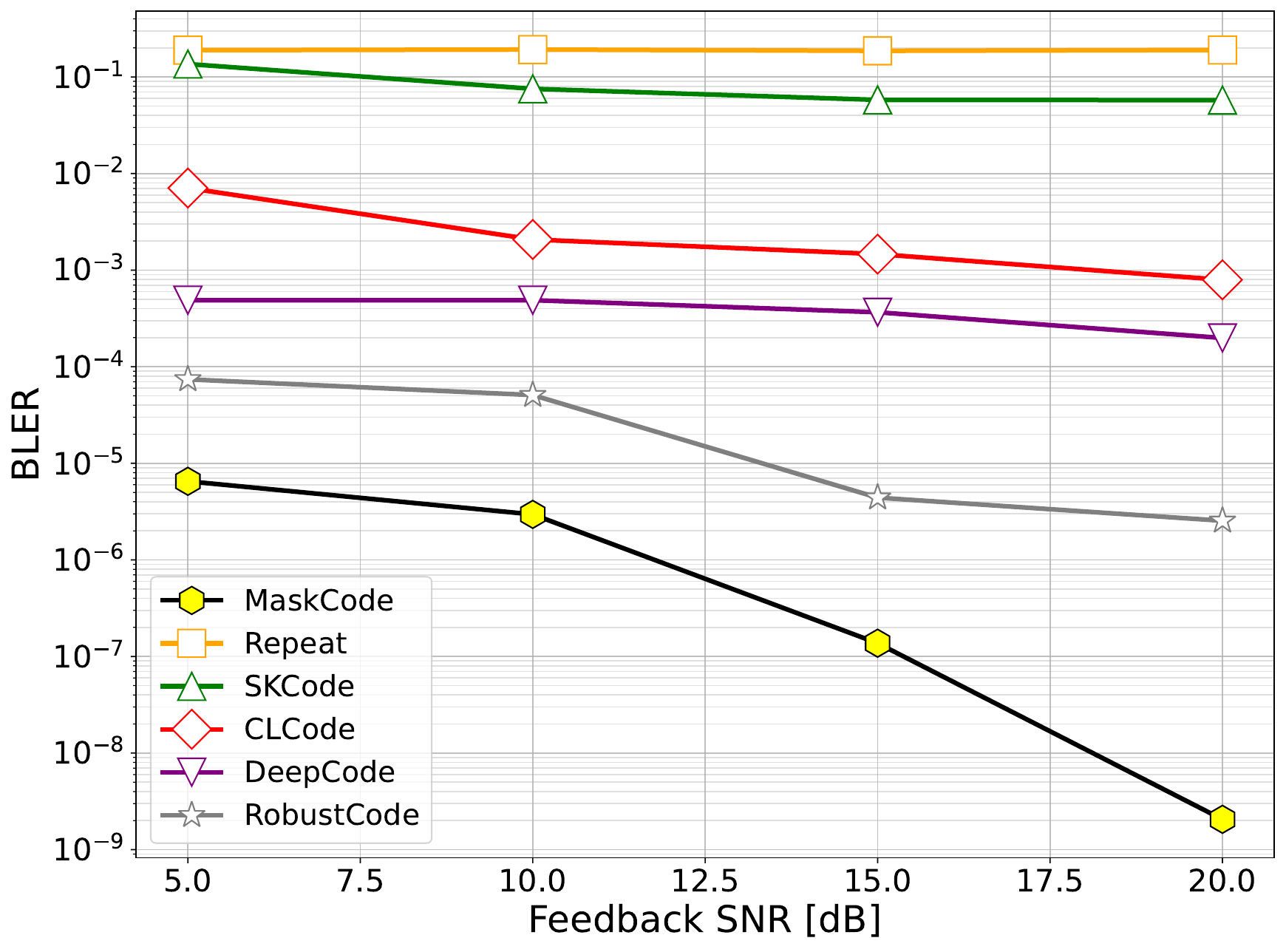}}
    \\
    \subfloat[LDPC(49,24) Outer Code.\label{subfig:fb_snr_sweep_ldpc}]{\includegraphics[width=0.9\linewidth]{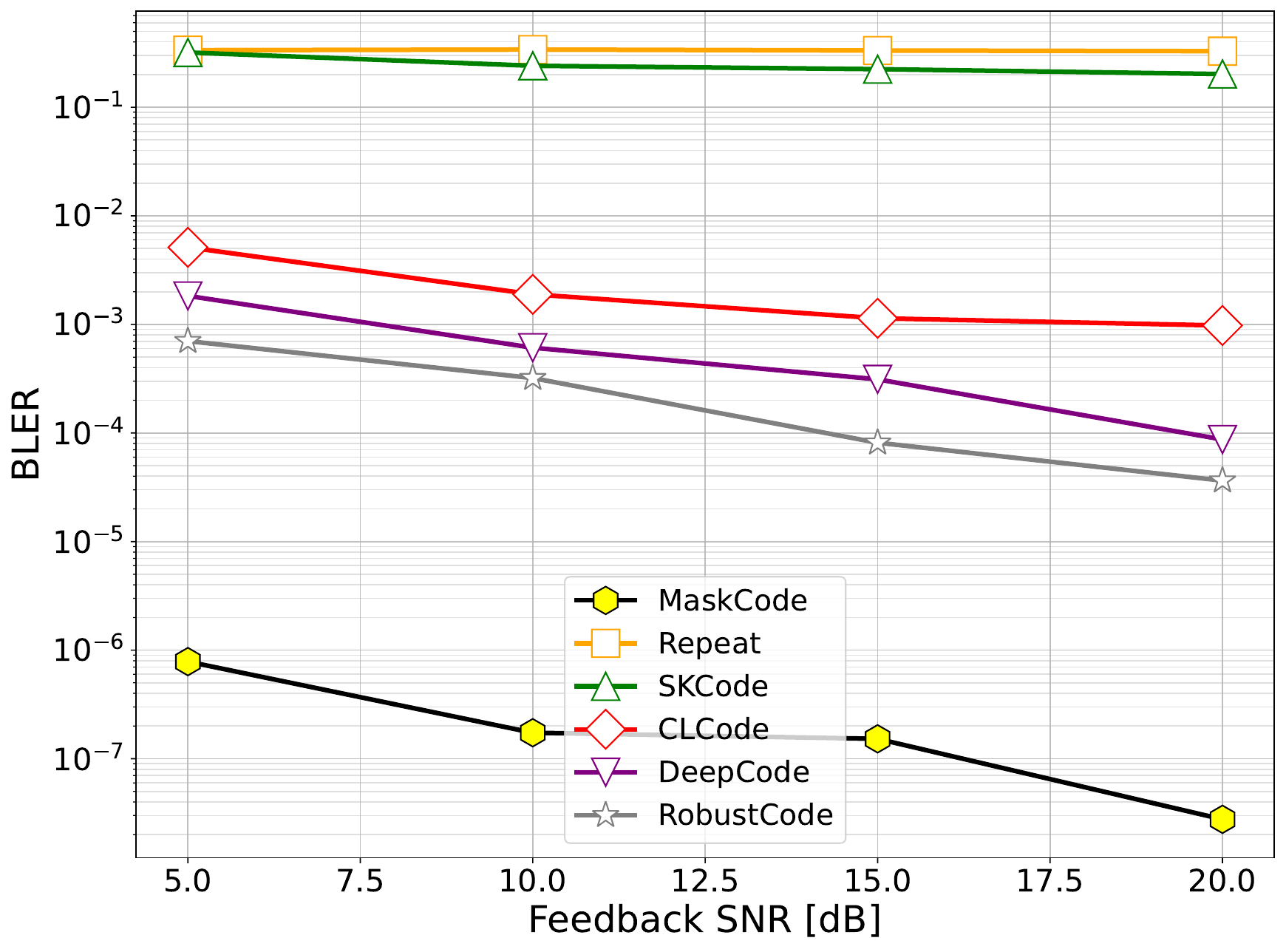}}
    \caption{BLER evaluation of \MaskCode~and baselines under various channel feedback noise levels. The forward SNR is -2 dB in \protect\subref*{subfig:fb_snr_sweep_bch} and -3 dB in \protect\subref*{subfig:fb_snr_sweep_ldpc}.\label{fig:fb_snr_sweep}}
\end{figure}

In this subsection, we evaluate the robustness of \MaskCode~to varying feedback noise levels in \cref{fig:fb_snr_sweep}. 
In the figure, \MaskCode~consistently outperforms all baselines across the entire feedback SNR regime for both outer codes. 
Notably, while the baselines exhibit a performance plateau around a BLER of $10^{-4}$, \MaskCode~demonstrates a continuous BLER decay, reaching as low as $10^{-9}$.

\subsection{Various Numbers of Layers}
\label{subsec:various_numbers_of_layers}
\begin{table}[t]
    \centering
    \caption{BLER performance for various numbers of layers in the \MaskCode. \label{tab:varous_numbers_of_layers} }
    
    \adjustbox{width=1\linewidth}{
    \begin{tabular}{ccccc}
    \toprule
        
        \multirow{2}{*}{Outer Code} & \multirow{2}{*}{Methods} & \multicolumn{3}{c}{BLER}
        \\ \cmidrule(lr){3-5}
        & & \makecell{$@$SNR=-3dB} & \makecell{$@$SNR=-4dB} & \makecell{$@$SNR=-5dB} \\ 
        \midrule
        \multirow{2}{*}{BCH(31,16)} & \MaskCode-S &   9.25$\times10^{-6}$ & 1.74$\times10^{-4}$ & 1.71$\times10^{-3}$ \\
        & \MaskCode-L &  7.04$\times10^{-7}$ & 5.09$\times10^{-5}$ & 1.10$\times10^{-3}$ \\
    \bottomrule
    \end{tabular}
    }
\end{table}

We examine the effect of model depth by comparing two configurations of \MaskCode~on the BCH(31,16) outer code:
\begin{itemize}
    \item \MaskCode-S: the standard 2-layer configuration used in all previous results.
    \item \MaskCode-L: a deeper variant of \MaskCode, with 4 layers.
\end{itemize}
Table \ref{tab:varous_numbers_of_layers} summarizes the BLER performance of the two configurations across varying feedforward SNR values. 
\MaskCode-L significantly outperforms \MaskCode-S across all SNR values, demonstrating that model depth is an important factor in concatenated coding performance.
This finding establishes a clear \textit{trade-off between performance and model complexity}.

\subsection{Computational Complexity}

\begin{table}[t]
    \centering
    \caption{Computational Complexity, multiply-accumulates (MACs) and wall-clock time}
    \adjustbox{width=.85\linewidth}{
    \begin{tabular}{cccc}
    \toprule
        Methods & Parameters & MAC & Codewords/sec \\ 
        \midrule
        \MaskCode~\textbf{(Ours)} & 5.46$\times10^5$ & 4.02$\times10^5$ & 175.44 \\ 
        AttentionCode & 1.26$\times10^5$ & 8.43$\times10^5$ & 17.95 \\ 
        GBAF & 1.55$\times10^5$ & 4.13$\times10^5$ & 65.17 \\ 
        \midrule
        DeepCode & 1.47$\times10^5$ & 4.55$\times10^6$ & 713.29 \\ 
        RobustCode & 1.02$\times10^5$ & 8.33$\times10^6$ & 739.13\\
    \bottomrule
    \end{tabular}
    }
    \label{tab:complexity}
\end{table}

Lastly, we evaluate the computational complexity and practical latency of \MaskCode~against the baselines, with results presented in \cref{tab:complexity}. 
We report three metrics: the number of parameters, multiply-accumulates (MACs), and wall-clock throughput (codewords/sec).

\paragraph*{Parameter counts and MAC}
The baselines fall into two distinct groups. 
The RNN-based methods (DeepCode and RobustCode) have a small parameter count but an extremely high MAC count (\eg, $4.55\times10^6$ for DeepCode), an order of magnitude higher than attention-based methods. 

\paragraph*{Throughput}

Despite its larger parameter count, \MaskCode~achieves the lowest MAC count among Transformer-based methods ($4.02\times10^5$), comparable to GBAF ($4.13\times10^5$) and significantly lower than AttentionCode ($8.43\times10^5$). More importantly, \MaskCode~achieves the highest throughput among Transformer-based methods at 175.44 codewords/sec, approximately 2.7 times faster than GBAF and nearly 10 times faster than AttentionCode.

\section{Conclusion}

In this paper, we propose \MaskCode, a novel Transformer-based inner feedback code designed for concatenated coding. 
The main insight motivating our design is that existing DNN-based feedback codes are optimized independently of the outer decoder's structure, misallocating feedback resources toward error patterns that fall within the outer ECC's correction capability. 
To address this, \MaskCode~incorporates the structural constraint of the outer linear block code into the inner feedback encoder via two synergistic mechanisms that explicitly: 1) a code-aware attention mask derived from the Tanner graph and 2)  a soft syndrome-based input design that informs the encoder about potential parity constraint violations.
While end-to-end training allows the inner code to account for the outer decoder's behavior, we empirically show that \MaskCode~achieves its best performance without any BP iterations in training, as the structural constraint already internalized in the architecture subsumes the benefit that end-to-end training would otherwise provide, while avoiding the gradient instability introduced by backpropagation through BP iterations.
Extensive evaluations on both BCH and LDPC outer codes demonstrate that \MaskCode~consistently outperforms all baselines across various experimental settings.

\bibliographystyle{IEEEtran}
\bibliography{main}

\appendices

\section{Quantized Feedback Scenario}

\begin{figure}
    \centering
    \includegraphics[width=0.99\linewidth]{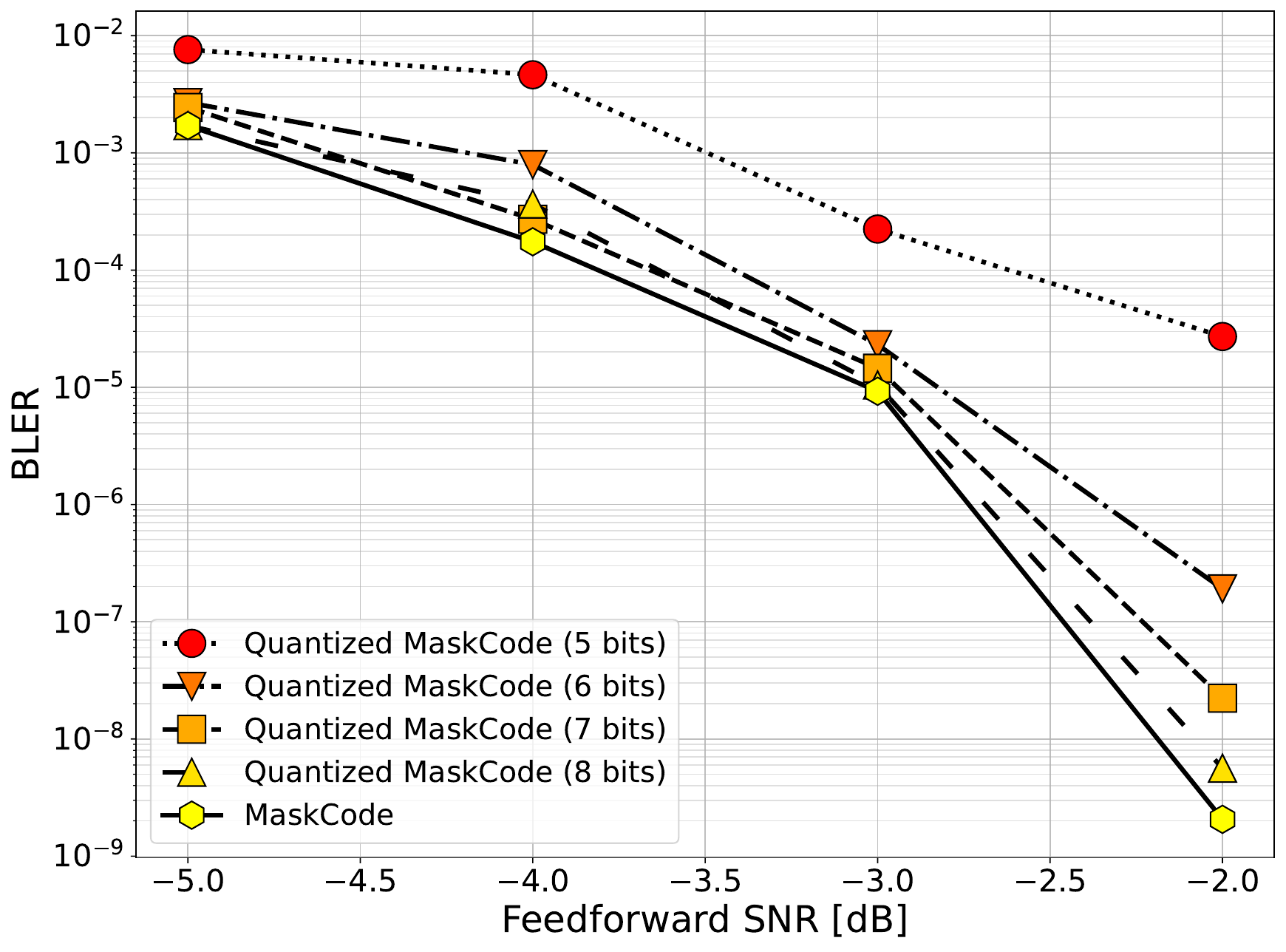}
    \caption{BLER performance of \MaskCode~under various feedback quantization levels. The \MaskCode~with 8-bit feedback quantization has a comparable BLER to that with full precision (32 bit). The outer code is the BCH(31,16) code.}
    \label{fig:quantization_bch}
\end{figure}

To evaluate the practical feasibility of \MaskCode, we investigate its performance under quantized feedback in \cref{fig:quantization_bch}, where the feedback signal is quantized to varying resolutions from 5 to 32 bits, with the BCH(31,16) outer code and feedforward SNR of $-2$ dB.
While low-resolution quantization of 5 or 6 bits leads to a noticeable BLER degradation, the performance improves consistently as the bit-depth increases.
At 8-bit quantization, the BLER of \MaskCode~converges to that of the full-precision (32-bit) case, demonstrating that \MaskCode~is robust to quantization noise and can be effectively deployed in practical hardware environments with limited feedback link capacity.

\begin{figure*}[!h]
    \centering
    \includegraphics[width=0.9\linewidth]{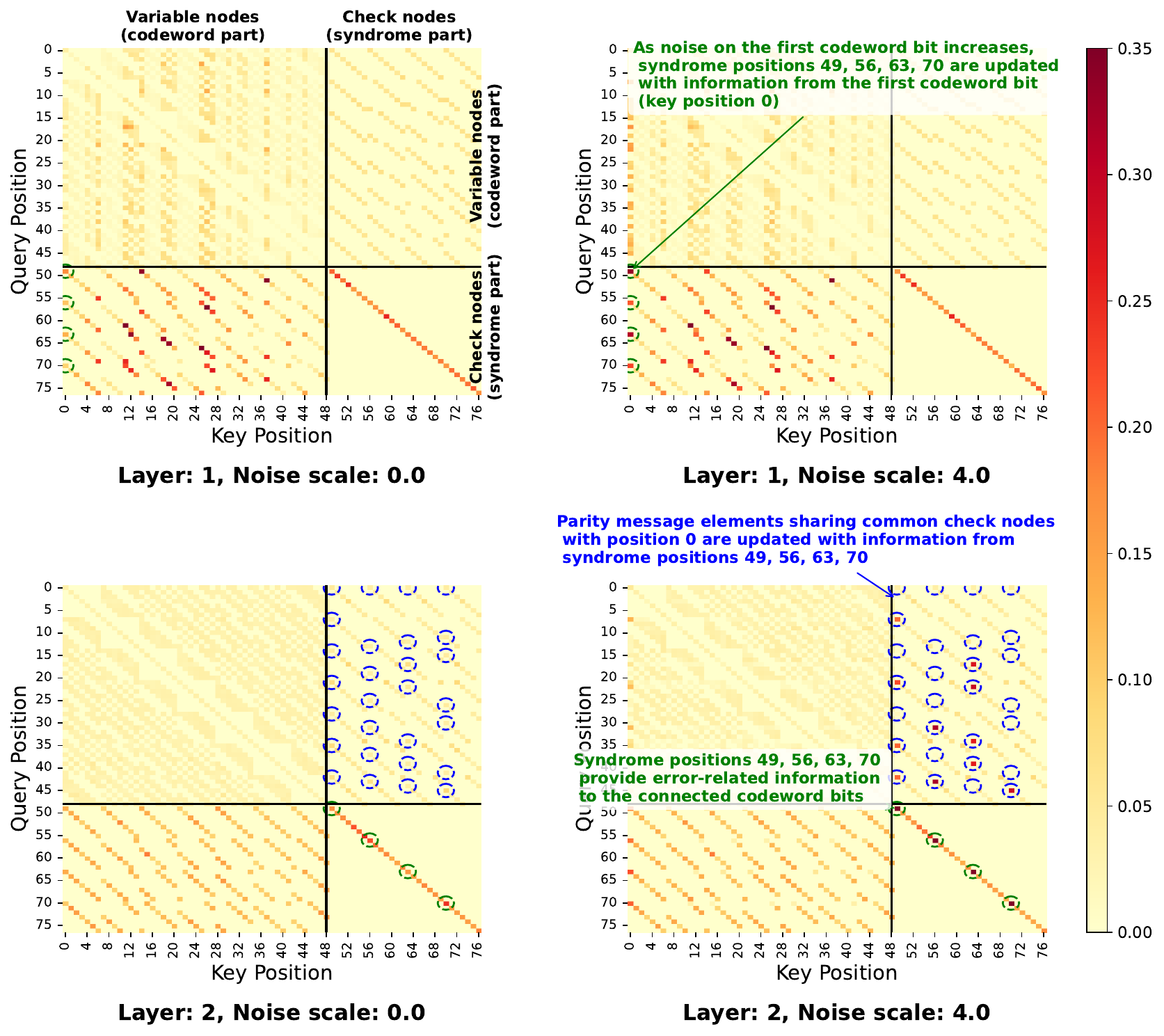}
    \caption{Illustration of self-attention maps at different layers of the \MaskCode~encoder layer in the second phase; the outer code is the LDPC(49,24) code. 
    Self-attention maps are shown for two noise levels (scale = 0.0 and 4.0) injected into the first element of $\mathbf{x}_1$, while noises on all other elements are set to zero.}
    \label{fig:attention_layer2_bch}
\end{figure*}

\section{Attention Mask Analysis}

Here, we analyze the self-attention maps to investigate the internal operations of the \MaskCode~encoder.
In \cref{fig:attention_layer2_bch}, we specifically examine the first and second layers of the encoder during the second phase ($t=2$) with the LDPC(49,24) code. 
In the second phase, the encoder generates the parity message $\mathbf{x}_2$ after receiving the initial feedback.
To observe how the model reacts to specific errors, we inject controlled noise into the first-phase message $\mathbf{x}_1$. 
More specifically, the noise corresponding to the first element of $\mathbf{x}_1$ is varied, while the noise injected into all other elements is set to zero. 
According to the parity-check matrix $\mathbf{H}$, the syndrome positions associated with the first codeword bit ($[\mathbf{H}]_{i,1}=1$) are at indices 49, 56, 63, and 70.

\paragraph*{First layer (Error recognition)}
The attention maps of the first layer are shown in the first row of \cref{fig:attention_layer2_bch}. 
Comparing the two noise levels, the syndrome positions 49, 56, 63, and 70 (highlighted in green) exhibit progressively stronger attention toward the first codeword bit as noise increases, indicating that these syndrome positions are updated with information from the corrupted bit.
This indicates that the first layer actively identifies potential parity constraint violations by focusing on the specific codeword bit corrupted by noise.

\paragraph*{Second layer (Structured refinement)} 
Building on the error recognition in the first layer, the second layer further processes the syndrome terms at positions 49, 56, 63, and 70, which now carry error-related information identified in the first layer, as depicted in the second row of \cref{fig:attention_layer2_bch}.
More specifically, the parity message elements at positions sharing common check nodes with the first codeword bit (highlighted in blue) are refined using information from syndrome positions 49, 56, 63, and 70.
This layer-by-layer progression demonstrates that \MaskCode~does not merely treat bits independently but instead performs a structured message-passing-like operation to reinforce bits that the outer decoder might find vulnerable.
Thus, our proposed synergistic design---combining the \textit{code-aware mask} and \textit{syndrome-based input}---successfully guides the Transformer to adhere to the structure of outer ECCs.

\end{document}